\documentclass[%
 reprint,
showpacs,preprintnumbers,
amsmath,amssymb,
aip,
jcp,
]{revtex4-1}

\usepackage{graphicx} 
\usepackage{dcolumn} 
\usepackage{bm} 
\usepackage{color}
\usepackage[mathlines]{lineno}

\begin{document}

\preprint{AIP/123-QED}

\title{Another resolution of the configurational entropy paradox as applied to hard spheres}%

\author{Vasili Baranau}
 \email{vasili.baranov@gmail.com}
\author{Ulrich Tallarek}
\affiliation{%
Department of Chemistry, Philipps-Universit\"{a}t Marburg, Hans-Meerwein-Strasse 4, 35032 Marburg, Germany
}%

\date{\today}

\begin{abstract}

Recently, Ozawa and Berthier [M. Ozawa and L. Berthier, \textit{J. Chem. Phys.}, 2017, \textbf{146}, 014502] 
studied the configurational and vibrational entropies $S_{\text{conf}}$ and $S_{\text{vib}}$ from 
the relation $S_{\text{tot}} = S_{\text{conf}} + S_{\text{vib}}$ for polydisperse mixtures of spheres.
They noticed that because the total entropy per particle $S_{\text{tot}} / N$ shall contain 
the mixing entropy per particle $k_B s_{\text{mix}}$ and $S_{\text{vib}} / N$ shall not, 
the configurational entropy per particle $S_{\text{conf}} / N$ shall diverge in the thermodynamic limit 
for continuous polydispersity due to the diverging $s_{\text{mix}}$. 
They also provided a resolution for this paradox and related problems---it relies on a careful redefining of $S_{\text{conf}}$ and $S_{\text{vib}}$.
Here, we note that the relation $S_{\text{tot}} = S_{\text{conf}} + S_{\text{vib}}$ is essentially a geometric relation
in the phase space and shall hold without redefining $S_{\text{conf}}$ and $S_{\text{vib}}$.
We also note that $S_{\text{tot}} / N$ diverges with $N \to \infty$ with 
continuous polydispersity as well. The usual way to avoid this and other difficulties with $S_{\text{tot}} / N$ 
is to work with the excess entropy $\Delta S_{\text{tot}}$ (relative to the ideal gas of the same polydispersity).
Speedy applied this approach to the relation above in [R. J. Speedy, \textit{Mol. Phys.}, 1998, \textbf{95}, 169] 
and wrote this relation as $\Delta S_{\text{tot}} = S_{\text{conf}} + \Delta S_{\text{vib}}$.
This form has flaws as well, because $S_{\text{vib}} / N$ does not contain the $k_B s_{\text{mix}}$ term and the latter 
is introduced into $\Delta S_{\text{vib}} / N$ instead.
Here, we suggest that this relation shall actually be 
written as $\Delta S_{\text{tot}} = \Delta_c S_{\text{conf}} + \Delta_v S_{\text{vib}}$,
where $\Delta = \Delta_c + \Delta_v$ while $\Delta_c S_{\text{conf}} = S_{\text{conf}} - k_B N s_{\text{mix}}$ and 
$\Delta_v S_{\text{vib}} = S_{\text{vib}} - k_B N \left[ 1 + \ln \left( \frac{V}{\Lambda^d N} \right) + \frac{U}{N k_B T} \right]$ 
with $N$, $V$, $T$, $U$, $d$, and $\Lambda$ standing for the number of particles, volume, temperature, internal energy, dimensionality, and de Broglie wavelength, respectively.
In this form, all the terms per particle are always finite for $N \to \infty$ and 
continuous when introducing a small polydispersity to a monodisperse system.
We also suggest that the Adam--Gibbs and related relations shall in fact contain $\Delta_c S_{\text{conf}} / N$ instead of $S_{\text{conf}} / N$.

\end{abstract}

\maketitle

\section{\label{sec:Introduction} Introduction}

\subsection{\label{subsec:Paradox} The paradox}

When studying glasses and glass-like systems, like colloids, 
it is typical to separate the total entropy of a system into the 
configurational and vibrational parts:\cite{speedy_hard_1998, 
stillinger_kauzmann_2001, sastry_relationship_2001, angelani_configurational_2007, donev_configurational_2007, 
foffi_short_2008, starr_relationship_2013} 
$S_{\text{tot}} = S_{\text{conf}} + S_{\text{vib}}$.
Here, $S_{\text{conf}}$ enumerates the states around which the system vibrates and $S_{\text{vib}}$ corresponds to 
the average volume in the phase space for vibrations around a single such state---i.e., vibrations in a basin of attraction of a configuration.

The total entropy in the canonical ensemble $S_{\text{tot}}$ is expressed in the standard way through the Helmholtz free energy $A_{\text{tot}}$, 
internal energy $U$ and temperature $T$ as 
$A_{\text{tot}} = U - T S_{\text{tot}}$, while $A_{\text{tot}}$ is 
expressed through the total partition function $Z_{\text{tot}}$ as $A_{\text{tot}} = -k_B T \ln Z_{\text{tot}}$. 
Thus, $S_{\text{tot}} = U / T + k_B \ln Z_{\text{tot}}$. In turn, $Z_{\text{tot}}$ is expressed through 
the configurational integral as $Z_{\text{tot}} = \frac{1}{\Pi_{t=1}^M N_t!} \frac{1}{\Lambda^{d N}} \int_{V^N} e^{-U_N(\vec{r})/k_B T} \text{d} \vec{r}$,
where $\Lambda$ denotes the de Broglie thermal wavelength $\Lambda = h / \sqrt{2 \pi m k_B T}$ (given that all particles have the same mass $m$).
The term $\frac{1}{\Pi_t N_t!}$ accounts for ``indistinguishability'' of constituent particles: there are in total $N$ 
particles with $M$ particle species.\cite{asenjo_numerical_2014, martiniani_turning_2016} 
It does not necessarily stem from quantum indistinguishability, but rather from our choice 
which particles we consider interchangeable to still be able to say 
that switching a pair of particles leaves the configuration 
unchanged.\cite{frenkel_why_2014, meng_free_2010}
For example, in colloids of sphere-like particles it is typical to consider particles with the same radius to be of the same type, 
though surface features apparently can allow to distinguish any pair of particles. This ``colloidal'' indistinguishability term
is needed to prevent the Gibbs paradox and define entropy in a reasonable way 
(so that the entropy is extensive).\cite{asenjo_numerical_2014, frenkel_why_2014, martiniani_turning_2016}
For hard spheres, $U_N(\vec{r}) = 0$ if there are no intersections between particles and $U_N(\vec{r}) = \infty$ otherwise.

If we consider entropy per particle $S_{\text{tot}} / N k_B$ (in units of $k_B$), it contains the term 
$\frac{1}{N} \ln \left( \frac{1}{\Pi_t N_t!} \right)$. With the help of the Stirling approximation $\ln(N) = N \ln(N) - N$, one 
obtains for the thermodynamic limit $N \to \infty$\cite{ozawa_does_2017}
\begin{equation}
    \frac{1}{N} \ln \left( \frac{1}{\Pi_t N_t!} \right) = 1 + s_{\text{mix}} - \ln(N),
    \label{eq:PolydiperseStirling}
\end{equation}
where $s_{\text{mix}} = - \sum_t^M \frac{N_t}{N} \ln \frac{N_t}{N}$ is the mixing entropy per particle (in units of $k_B$) 
or the information entropy of 
the particle type distribution. This quantity diverges in the thermodynamic limit in the case of a continuous particle type 
distribution\cite{baranau_chemical_2016, ozawa_does_2017}---
for example, if spherical colloidal particles have a continuous radii distribution $f(r)$. 
Indeed, if we discretize the distribution with the step $\delta$, then $N_t = N f(r_t) \delta$ in the limit $\delta \to 0$ and 
\begin{equation}
    s_{\text{mix}} = - \int f(r) \ln(f(r)) \text{d} r - \ln(\delta),~ \delta \to 0.
    \label{eq:MixingEntropyDivergence}
\end{equation}
The mixing entropy per particle diverges due to the 
diverging $\ln(\delta)$ term.
In information theory, it is typical to work with the differential entropy when dealing with continuous probability 
distributions,\cite{stone_information_2015, lazo_entropy_1978}
where the differential entropy is the right hand side of
Eq. (\ref{eq:MixingEntropyDivergence}) without the $- \ln(\delta)$ term,
$s_{\text{dif}} = - \int f(r) \ln(f(r)) \text{d} r$. 
Given that for a uniform distribution in the interval $[a, b]$ 
$s_{\text{dif}} = \ln(b-a)$,\cite{lazo_entropy_1978} 
$s_{\text{dif}}$ of an arbitrary function is 
its information entropy ($s_{\text{mix}}$) with respect to 
the uniform distribution in a unit interval $[0, 1]$.
The $\ln(N)$ term in Eq. (\ref{eq:PolydiperseStirling}) does not pose a problem in the thermodynamic limit, because it is in fact incorporated 
into the $\ln(V / \Lambda^d N)$ term in $S_{\text{tot}} / N k_B$ (\textit{cf}. Eq. (\ref{eq:TotalEntropy}) below).
Of course, to ensure applicability of the Stirling approximation, 
limits $N \to \infty$ and 
$\delta \to 0$ shall be taken carefully: we shall always ensure that 
$N \delta \min f(r) \gg 1$. If $f$ has an infinite support, 
\textit{e.g.} $[0, +\infty)$, we have to sample radii from 
an interval $[0, R(N))$ (where $R \to \infty$ with $N \to \infty$), 
still imposing 
$N \delta \min \limits_{[0, R)} f(r) \gg 1$. 
To ensure that the Stirling approximation becomes precise in 
the thermodynamic limit, we can choose a certain scaling for 
$a = N \delta \min f(r)$ as well.
For example, for $f(r)$ with a finite support 
(and a fixed $f_{\text{min}} = \min f(r)$), 
we can select $a(N) = \sqrt{N}$ or in general $a(N) = N^\gamma$, 
$\gamma < 1$, so that 
$\delta = N^{\gamma - 1} f_{\text{min}}^{-1} \to 0$ with $N \to \infty$.

Now, if we restrict the phase space only to a certain 
basin of attraction $V_{\text{basin}}$, 
we can write similar to $S_{\text{tot}}$ $S_{\text{vib}} = U / T + k_B \ln Z_{\text{vib}}$. $Z_{\text{vib}}$ 
is often written as $Z_{\text{vib}} = \frac{1}{\Pi_t N_t!} \frac{1}{\Lambda^{d N}} \int_{\text{basin}} e^{-U_N(\vec{r})/k_B T} \text{d} \vec{r}$.
But for every basin of attraction there are exactly $\Pi_t N_t!$ equivalent basins due to particle permutations. 
Because the $\Pi_t N_t!$ terms are compensated, $Z_{\text{vib}}$ shall actually be expressed as 
$Z_{\text{vib}} = \frac{1}{\Lambda^{d N}} \int_{\text{basin}} e^{-U_N(\vec{r})/k_B T} \text{d} \vec{r}$.\cite{ozawa_does_2017}
This fact was realized for monodisperse systems long ago.\cite{stillinger_limiting_1969, speedy_entropy_1993, speedy_pressure_1998}

If we now consider the equation $S_{\text{tot}}/k_B N = S_{\text{conf}}/k_B N + S_{\text{vib}}/k_B N$, 
$S_{\text{tot}}/k_B N$ contains the $s_{\text{mix}}$ term 
while $S_{\text{vib}}/k_B N$ does not. Ozawa and Berthier\cite{ozawa_does_2017} pointed out that there are several problems with this.
Firstly, it means that $S_{\text{conf}}/k_B N$ shall diverge in the thermodynamic limit for a continuous 
particle type distribution with diverging $s_{\text{mix}}$.
Secondly, if we take a colloid with spherical particles of equal size and introduce a slight polydispersity, 
$s_{\text{mix}}$ will exhibit a jump (from zero to a non-zero value, e.g. $\ln 2$ in the case of a $50:50$ binary mixture, 
however similar particle radii are).
$S_{\text{conf}}/k_B N$ will exhibit the same jump. These are the two basic problems that constitute the paradox.

\subsection{\label{subsec:ResolutionOfOzawaAndBerthier} Resolution of Ozawa and Berthier}

Ozawa and Berthier suggested\cite{ozawa_does_2017} to carefully redefine entropies: roughly, to ``merge'' (besides the $\Pi_t N_t!$ merging) those basins 
that have high overlap, \textit{i.e.}, that look sufficiently similar to each other due to similar particle types and 
constituent configurations. This procedure decreases the effective number of configurations 
around which the system is considered to vibrate and compensates the jump in $S_{\text{conf}}/k_B N$ when particles are 
made only slightly different from each other. In other words, $S_{\text{conf}}/k_B N$ will behave continuously 
when particles are made only slightly different from each other. 
This procedure also essentially decreases the number of particle species and keeps $s_{\text{mix}}$ finite.
Ref.\cite{ozawa_does_2017} assumes that original basins (before redefinitions)
are defined in some sort of a free energy landscape\cite{charbonneau_fractal_2014, charbonneau_glass_2017} 
(\textit{e.g.}, emerging from the density-functional theory, 
where a state is a particular spatial density profile).

\subsection{\label{subsec:RemainingProblems} Motivations for another resolution}

The resolution of Ozawa and Berthier is perfectly valid, 
but it relies on ``merging'' the basins with high overlap and is thus 
not applicable if for some reason 
we do not want to do any redefinition or ``merging'' of basins 
(except for the $\Pi_t N_t!$ merging), 
even if they have high overlaps.
One popular definition of basins uses steepest descents in the potential energy 
landscape (PEL).\cite{stillinger_systematic_1964, stillinger_topographic_1995, 
debenedetti_supercooled_2001, torquato_robust_2010}
$S_{\text{conf}}$ is then defined\cite{donev_configurational_2007, ashwin_calculations_2012, asenjo_numerical_2014, martiniani_turning_2016} 
through the number of local PEL minima or 
inherent structures\cite{stillinger_systematic_1964, stillinger_topographic_1995, torquato_robust_2010} 
(that still have to be merged due to the ``colloidal'' indistinguishability of particles). 
The vibrational entropy is then defined through basins of attraction in the PEL. 
For hard spheres, one has to use a pseudo-PEL,\cite{torquato_robust_2010} where 
inherent structures correspond\cite{torquato_robust_2010, baranau_random_close_2014, zinchenko_algorithm_1994} to jammed configurations.
These definitions are mathematically precise and allow splitting the phase space into basins even for the ideal gas.
For simple systems, decomposition of the phase space into basins can be done numerically by doing steepest 
descents from many starting configurations.\cite{ashwin_calculations_2012, asenjo_visualizing_2013} 
The resolution of Ozawa and Berthier is not applicable to these definitions 
(\textit{i.e.}, if we require keeping the basins from these definitions unchanged), 
but the relation $S_{\text{tot}} = S_{\text{conf}} + S_{\text{vib}}$ shall be valid, because it is essentially a geometrical 
relation that tells us how the phase or configuration space is split into volumes around some points.
It shall be valid for an arbitrary decomposition of the configuration space into basins, the only requirement being 
the saddle point approximation.\cite{speedy_hard_1998}

\subsection{\label{subsec:OtherResolutions} Other previous resolutions}

We mentioned that $S_{\text{tot}}/N k_B$ contains $s_{\text{mix}}$.
It means that $S_{\text{tot}}/N k_B$ alone has all the problems that Ozawa and Berthier were solving: (i) discontinuity 
with introduction of a small polydispersity into a monodisperse system and (ii) divergence with a continuous particle type distribution.
This is not a severe problem, because in experiments only entropy differences or entropy derivatives matter 
($S_{\text{conf}}$ is such a difference).
But whenever an equation containing $S_{\text{tot}}/N k_B$ is valid in the polydisperse case, there shall be other terms that 
cancel $s_{\text{mix}}$ exactly. Thus, it makes sense to write such equations through non-diverging terms, 
when all equivalently divergent terms are omitted.
Hence, a lot of well-known papers on hard-sphere fluids and glasses, including the classical ones by Carnahan and Starling, 
work solely with the excess entropy $\Delta S_{\text{tot}}$ (with respect to the ideal gas of the corresponding particle size distribution), 
which does not contain the unpleasant term $s_{\text{mix}}$.\cite{carnahan_thermodynamic_1970, mansoori_equilibrium_1971, adams_chemical_1974, 
speedy_entropy_1993, speedy_pressure_1998, speedy_hard_1998, donev_calculating_2007} 
For example, the equation that connects $S_{\text{tot}}$, the chemical potential $\mu$ and reduced 
pressure $Z = p V / N k_B T$ in equilibrium polydisperse hard-sphere systems is written as 
$\Delta S_{\text{tot}} / N k_B = Z - 1 - \langle \Delta \mu \rangle / k_B T$.\cite{carnahan_thermodynamic_1970, 
mansoori_equilibrium_1971, adams_chemical_1974, baranau_chemical_2016}

Speedy used the same approach of working with excess quantities when studying glassy systems of hard spheres and the relation 
$S_{\text{tot}} = S_{\text{conf}} + S_{\text{vib}}$ as early as in $1998$.\cite{speedy_hard_1998} He wrote this relation in the form 
$\Delta S_{\text{tot}} = S_{\text{conf}} + \Delta S_{\text{vib}}$. As we explained above, this form seems to be not the complete 
resolution of the problems with these quantities as well, because $S_{\text{vib}} / N k_B$ does not actually contain the $s_{\text{mix}}$
term, so it is rather introduced into the $\Delta S_{\text{vib}} / N k_B$ instead of being removed.

\subsection{\label{subsec:OurResoultion} Overview of our resolution}

We believe that the general strategy of working with excess quantities is the one to follow, 
but the approach of Speedy\cite{speedy_hard_1998} shall be slightly revised.
We show that instead of writing $\Delta S_{\text{tot}} = S_{\text{conf}} + \Delta S_{\text{vib}}$ one has to write
$\Delta S_{\text{tot}} = \Delta_c S_{\text{conf}} + \Delta_v S_{\text{vib}}$, 
where $\Delta_c$ is an operator that subtracts $k_B N s_{\text{mix}}$ 
and $\Delta_v$ is an operator that subtracts 
$k_B N [1 + \ln \left( \frac{V}{\Lambda^d N} \right) + \frac{U}{N k_B T}]$. Together, they are equivalent to $\Delta$ 
(in the operator sense, if all the operators are applied to a common variable, 
$\Delta = \Delta_c + \Delta_v$).

We start our discussion with an example of $1D$ hard ``spheres'' (rods) in a non-periodic system and 
then introduce the general case. We always assume that basins are defined through steepest descents in the (pseudo-)PEL 
and focus the discussion on hard spheres for simplicity.

Essentially, the main idea of the paper is how exactly we have to distribute the terms from the ideal gas entropy 
(subtracted from $S_{\text{tot}}$ to get $\Delta S_{\text{tot}}$) between $S_{\text{conf}}$ and $S_{\text{vib}}$.
They are distributed in an uneven way, similar to the relation 
$\Delta S_{\text{tot}} / N k_B = \Delta Z - \langle \Delta \mu \rangle / k_B T = Z - 1 - \langle \Delta \mu \rangle / k_B T$.

\subsection{\label{subsec:AdamGibbsIntroduction} The Adam--Gibbs 
and related relations}

As pointed out by one of the reviewers, this work would be 
incomplete without discussing which form of the configurational 
entropy shall be present in the Adam--Gibbs relation\cite{adam_temperature_1965, 
cavagna_supercooled_2009, starr_relationship_2013} 
(or in general any relation that connects the relaxation time 
of a system and $S_{\text{conf}}$, \textit{e.g.}, 
the one from the Random First Order Transition 
theory\cite{kirkpatrick_random_2015, cavagna_supercooled_2009, 
starr_relationship_2013, bouchaud_adam_2004}).
The Adam--Gibbs relation expresses the relaxation time of the system
$\tau_R$ through $S_{\text{conf}} / N$ as 
$\tau_R = \tau_0 \exp\left( \frac{A}{T S_{\text{conf}} / N} \right)$, 
where $\tau_0$ and $A$ are constants.
As explained above, $S_{\text{conf}} / N$ diverges for 
systems with continuous polydispersity for $N \to \infty$,
which means that $\tau_R \equiv 0$ for such systems according to the 
Adam--Gibbs theory.
It is also natural to assume that any relation 
for $\tau_R$ shall be continuous 
with respect to adding a slight polydispersity to a monodisperse system, 
because relaxation dynamics will remain almost unchanged. 
$S_{\text{conf}} / N$ has a jump in such a case, and $\tau_R$ from the
Adam--Gibbs relation will have it as well.
These two unphysical properties of the Adam--Gibbs relation indicate that 
it shall be amended.
We show below that our $\Delta_c S_{\text{conf}} / N$ is always finite 
and continuous and is thus a natural candidate
for the Adam--Gibbs and similar relations for the relaxation time. 
We discuss this aspect in more detail in 
Section \ref{sec:AdamGibbs}.

\section{\label{sec:OneDimensionalRods} Exactly solvable example: 1D rods in a non-periodic interval}

Let us look at first at a very simple 1D system: several rods (1D hard spheres) in a non-periodic interval.
For 3 rods, one can visualize the entire configuration space.

The phase space of 1D hard rods is never ergodic (but a single basin is), but we don't currently require ergodicity, 
because we only try to split the total configuration space 
into basins of attraction and look how different quantities scale with the number of particles.

\begin{figure}[t]
  \centering
  \includegraphics[width=0.5\textwidth]{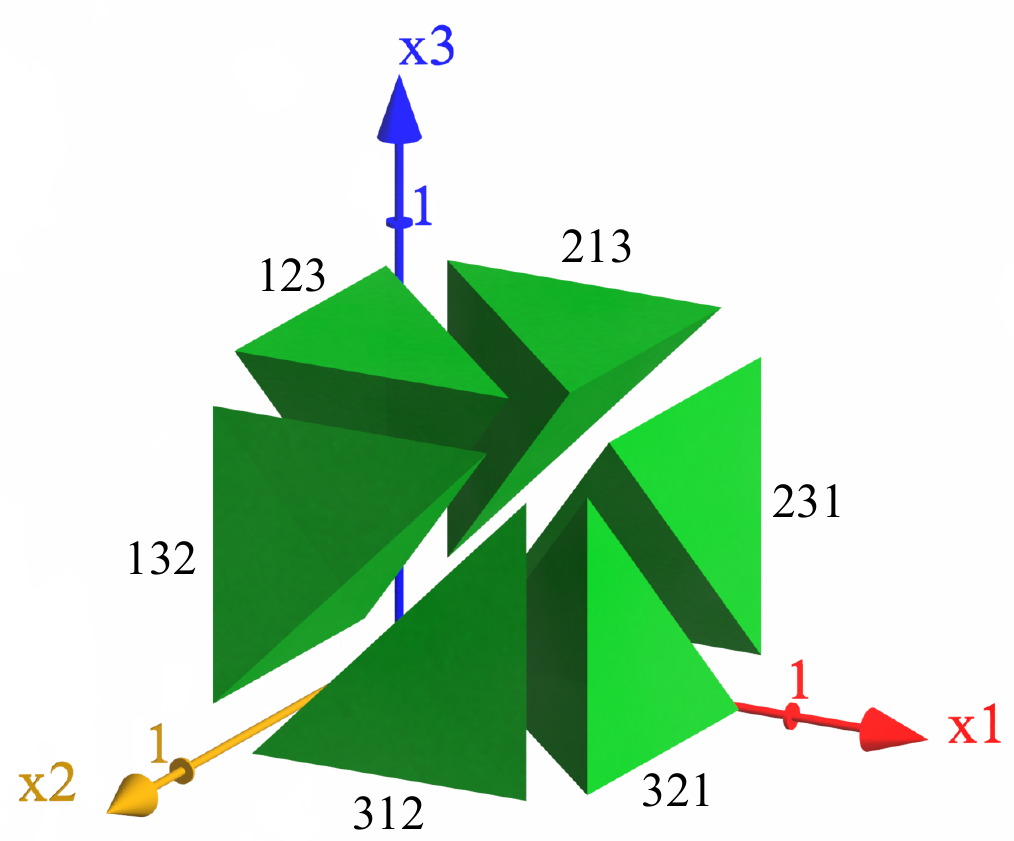}
  \caption{Configuration space of $N=3$ one-dimensional particles with non-periodic boundary conditions. 
  Numbers like ``231'' denote the order of particles in a corresponding jammed configuration.
  Reproduced with modifications from Ref.\cite{baranau_upper_2016} 
  }
  \label{fig:1D_PhaseSpace}
\end{figure}

The complete configuration space is presented in Fig. \ref{fig:1D_PhaseSpace}. 
It is a variant of Fig. 2 in Ref.\cite{stillinger_limiting_1969}, but these authors assumed periodic boundary conditions.

If all 3 particles can be distinguished from each other, 
there are 6 jammed configurations: 123 132 213 231 312 321 or $N!$ in the general case.
If the particles are monodisperse (of diameter A), 
it is essentially one configuration of indistinguishable particles: AAA.
Let us assume now that the particles are bidisperse, \#1---of type (diameter) A, \#2 and 3---of type B. 
If we treat them as indistinguishable, there are three jammed configurations: ABB, BAB, BBA or
$\frac{N!}{\Pi_t N_t!}$ in the general case.
The configurational entropy has a jump after switching to the bidisperse system, but this is natural,
while there are more distinct jammed configurations now. The total entropy has an equivalent jump.

Let us examine the vibrational, total, and configurational entropies of such a system. 
We assume for simplicity zero solid volume fraction of a system, 
$\varphi = V_{\text{spheres}} / V_{\text{box}} = 0$. 
The total volume of the configuration space is in this case simply $I_{\text{tot}} = V^N$. The total partition function is 
$Z_{\text{tot}} = \frac{1}{\Pi_t N_t!} \frac{1}{\Lambda^{dN}} V^N$, 
where we write $\Lambda^d$ with $d$ for dimensionality for the general case.

If all particles are treated as distinguishable, the number of jammed configurations is $N_J^{\text{dist}} = N! (= 6)$ and 
the average volume of a basin of attraction 
if all particles are treated as distinguishable 
is $I_{\text{vib}}^{\text{dist}} = \frac{V^N}{N!}$ 
(each green simplex in Fig. \ref{fig:1D_PhaseSpace}).
We can trivially write $I_{\text{tot}} = N_J^{\text{dist}} I_{\text{vib}}^{\text{dist}}$ or $V^N = N! \frac{V^N}{N!}$. 

If we treat particles as indistinguishable, 
we have to merge some jammed configurations to treat as a single one 
(divide $N_J^{\text{dist}}$ by $\Pi_t N_t!$)
and have to merge some basins of attraction to treat as a single one 
(multiply $I_{\text{vib}}^{\text{dist}}$ by $\Pi_t N_t!$).
The number of jammed configurations if particles are treated as 
indistinguishable is thus 
$N_J^{\text{ind}} = \frac{N!}{\Pi_t N_t!} (= 3)$.
The average volume of a basin of attraction 
if particles are treated as indistinguishable
is thus $I_{\text{vib}}^{\text{ind}} = 
\Pi_t N_t! \frac{V^N}{N!} (= \frac{V^N}{3})$. 
In Fig. \ref{fig:1D_PhaseSpace}, 
we have to merge the green tetrahedra in pairs.
We can write similar to the distinguishable case 
$I_{\text{tot}} = N_J^{\text{ind}} I_{\text{vib}}^{\text{ind}}$.
The vibrational partition function is then 
$Z_{\text{vib}}^{\text{ind}} = 
\frac{1}{\Pi_t N_t!} \frac{1}{\Lambda^{d N}} I_{\text{vib}}^{\text{ind}} =
\frac{1}{\Lambda^{d N}} \frac{V^N}{N!}$.
As mentioned in the introduction, the multiplication by the
number of permutations and the division by this number due 
to basin multiplicity always cancel out.

The total entropy per particle is expressed for our system as 
$S_{\text{tot}} / k_B N = U / N k_B T + \frac{1}{N} \ln Z_{\text{tot}} = 
U / N k_B T + 1 + s_{\text{mix}} + \ln \left( \frac{V}{\Lambda^d N} \right)$.
The vibrational entropy per particle (of indistinguishable particles) is expressed as 
$S_{\text{vib}} / k_B N \equiv S_{\text{vib}}^{\text{ind}} / k_B N = U / N k_B T + \frac{1}{N} \ln Z_{\text{vib}}^{\text{ind}} = 
U / N k_B T + 1 + \ln \left( \frac{V}{\Lambda^d N} \right)$.
The configurational entropy per particle (of indistinguishable particles) is by definition
$S_{\text{conf}} / k_B N \equiv S_{\text{conf}}^{\text{ind}} / k_B N = \frac{1}{N} \ln N_J^{\text{ind}} = \frac{1}{N} \ln \frac{N!}{\Pi_t N_t!} = s_{\text{mix}}$.

Naturally, these results conform to the equation 
$S_{\text{tot}} / k_B N = S_{\text{conf}} / k_B N + S_{\text{vib}} / k_B N$,
which is just another expression for the relation 
$I_{\text{tot}} = N_J^{\text{ind}} I_{\text{vib}}^{\text{ind}}$. 
The following result is surprising, though: the $s_{\text{mix}}$ term from 
$S_{\text{tot}} / k_B N$ is consumed on the right side of the equation by 
$S_{\text{conf}} / k_B N$  and the terms 
$1 + \ln \left( \frac{V}{\Lambda^d N} \right) + \frac{U}{N k_B T}$ 
from $S_{\text{tot}} / k_B N$ are consumed on the right side of 
the equation by $S_{\text{vib}} / k_B N$.
$S_{\text{vib}} / k_B N$ does not contain the mixing contribution, 
because $\frac{1}{\Pi_t N_t!}$ stemming from indistinguishability is 
exactly compensated by $\Pi_t N_t!$ stemming from basin multiplicity.
Thus, one can write
\begin{equation}
    \Delta S_{\text{tot}} = \Delta_c S_{\text{conf}} +
    \Delta_v S_{\text{vib}},
    \label{eq:CorrectExcessEntropies}
\end{equation}
where
\begin{equation}
\begin{aligned}
    &\frac{\Delta_c S_{\text{conf}}}{k_B N} = 
    \frac{S_{\text{conf}}}{k_B N} - s_{\text{mix}} \text{ and} \\
    &\frac{\Delta_v S_{\text{vib}}}{k_B N} = 
    \frac{S_{\text{vib}}}{k_B N} - 1 - \ln \left( \frac{V}{\Lambda^d N} \right) - \frac{U}{N k_B T}.
\end{aligned}
\end{equation}
All the terms in Eq. (\ref{eq:CorrectExcessEntropies}) 
if taken per particle are finite in the thermodynamic limit 
even for continuous particle size distributions 
and continuous with introduction of a small polydispersity to a monodisperse system.

\section{General theory, arbitrary $d$}

At first, we routinely derive the relation for the total entropy and demonstrate that 
$S_{\text{tot}} / k_B N$ contains $s_{\text{mix}} + \ln \left( \frac{V}{\Lambda^d N} \right)$ in the general case.

Then, we show that the relation 
$S_{\text{tot}} = S_{\text{conf}} + S_{\text{vib}}$ is truly a geometrical one 
and requires only a saddle point approximation. 
This approximation is actually exact in the thermodynamic limit.

Our next and main aim is then to show that the vibrational entropy per particle 
shall contain the term $\ln \left( \frac{V}{\Lambda^d N} \right)$, 
but not $s_{\text{mix}}$ in the general case as well, 
if ``colloidal'' indistinguishability of particles 
is treated carefully. It will mean that $s_{\text{mix}}$ is contained in $S_{\text{conf}} / k_B N$.

To investigate the volume of basins of attraction at arbitrary $\varphi$, 
we use a variant of thermodynamic integration.\cite{frenkel_new_1984, frenkel_understanding_2002, stillinger_kauzmann_2001, 
donev_calculating_2007, donev_configurational_2007, asenjo_numerical_2014} 

We use the same superscripts as before: ``dist'' as if all particles are 
distinguishable and ``ind'' as if particles of the same type (radius) 
are indistinguishable, implying that $S_{\text{conf}} \equiv S_{\text{conf}}^{\text{ind}}$ and 
$S_{\text{vib}} \equiv S_{\text{vib}}^{\text{ind}}$.

\subsection{Total entropy}

For the ideal gas, the integral 
$\int_{V^N} e^{-U_N(\vec{r})/k_B T} \text{d} \vec{r} = V^N$. 
Thus, the entropy of the ideal gas $S_{\text{tot}}^\circ$ is expressed with the help of 
$S_{\text{tot}}^\circ = U / T + k_B \ln Z_{\text{tot}}^\circ$ as
\begin{equation}
    \frac{S_{\text{tot}}^\circ}{N k_B} = \frac{U}{N k_B T} + 1 + s_{\text{mix}} 
    + \ln \left( \frac{V}{\Lambda^d N} \right),
    \label{eq:IdealGasEntropy}
\end{equation}
where $U / N k_B T = d/2$ ($= 3/2$ for $d=3$). 
We assume for the ideal gas the same relative 
particle radii distribution $f(r/ \langle r \rangle)$, but with $\langle r \rangle \to 0$.
The total entropy per particle can then be expressed as 
\begin{equation}
\begin{split}
    \frac{S_{\text{tot}}}{N k_B} =& \frac{S_{\text{tot}}^\circ}{N k_B} + \frac{\Delta S_{\text{tot}}}{N k_B} = \\
    =& \frac{U}{N k_B T} + 1 + s_{\text{mix}} + \ln \left( \frac{V}{\Lambda^d N} \right) + \frac{\Delta S_{\text{tot}}}{N k_B},
    \label{eq:TotalEntropy}
\end{split}
\end{equation}
where 
$\Delta S_{\text{tot}} / N k_B = (S_{\text{tot}} - S_{\text{tot}}^\circ)/N k_B = 
\frac{1}{N k_B} \ln \left( \frac{1}{V^N} \int_{V^N} e^{-U_N(\vec{r})/k_B T} \text{d} \vec{r} \right)$ 
is the excess entropy per particle in units of $k_B$. 
It can be shown through the definition of pressure 
$p = - (\partial A_{\text{tot}} / \partial V)_{N,T}$ that for hard spheres
\begin{equation}
    \frac{\Delta S_{\text{tot}}}{N k_B} = - \int \limits_{0}^\varphi \frac{Z(\varphi') - 1}{\varphi'} \text{d} \varphi',
    \label{eq:ExcessEntropy}
\end{equation}
where 
$\varphi = V_{\text{spheres}} / V_{\text{box}}$ is the solid volume fraction (packing density) and 
$Z(\varphi) = p V / N k_B T$ is the reduced pressure 
(\textit{cf}. Appendix \ref{subsubsec:TotalEntropyThroughPressures}).\cite{carnahan_thermodynamic_1970, adams_chemical_1974, speedy_hard_1998, 
donev_calculating_2007} Eq. (\ref{eq:ExcessEntropy}) can be regarded as 
a special case of thermodynamic integration.

The quantity $\Delta S_{\text{tot}} / N k_B$ does not share problems 
for $S_{\text{tot}}/N k_B$ mentioned above (divergence and discontinuity). 
Additionally, it does not contain the term $\ln \left( \frac{V}{\Lambda^d N} \right)$. Thus, 
$\Delta S_{\text{tot}} / N k_B$ depends on $\varphi$ only and does not depend on the temperature $T$ (which is indirectly present in $\Lambda$).

\subsection{Saddle point approximation: separation of the total entropy}

Even in the monodisperse case, the total volume of the configuration 
space $I_{\text{tot}}$ 
and the available volume of a basin of attraction 
$I_{\text{vib}}^{\text{ind}}$ depend on $\varphi$. 
The quantity $I_{\text{vib}}^{\text{ind}}$ 
is expressed (in the general---polydisperse---case) as 
$I_{\text{vib}}^{\text{ind}} = \Pi_t N_t! I_{\text{vib}}^{\text{dist}}$, where
$I_{\text{vib}}^{\text{dist}} = 
\int_{\text{basin}} e^{-U_N(\vec{r})/k_B T} \text{d} \vec{r}$.

With $d > 1$, there are jammed configurations at different 
densities.\cite{chaudhuri_jamming_2010, parisi_mean_field_2010, 
baranau_random_close_2014, baranau_jamming_2014, baranau_how_2015, 
coslovich_exploring_2017} 
Hence, there is a ``density of jamming densities'' 
$N_J^{\text{ind}} (N, \varphi_J)$.
We assume that properties of $I_{\text{vib}}^{\text{ind}}$ depend only on $N$, 
$\varphi$, and $\varphi_J$ (not on a particular basin of attraction), 
so we write $I_{\text{vib}}^{\text{ind}}(N, \varphi, \varphi_J)$.\cite{speedy_hard_1998, donev_configurational_2007}
Thus, we express the volume of the configuration space as
\begin{equation}
    I_{\text{tot}}(N, \varphi) = 
    \int \limits_\varphi^1 N_J^{\text{ind}} (N, \varphi_J) 
    I_{\text{vib}}^{\text{ind}}(N, \varphi, \varphi_J) \text{d} \varphi_J.
  \label{eq:SaddlePointStart}
\end{equation}

For any fixed $N$, $N_J^{\text{ind}} (N, \varphi_J)$ and 
$I_{\text{vib}}^{\text{ind}}(N, \varphi, \varphi_J)$ 
shall depend on $\varphi_J$ as follows:
$N_J^{\text{ind}}$ shall decrease rapidly with the increase of $\varphi_J$ 
(and more rapidly with larger $N$), as indicated by numerous results 
on the configurational entropy and relaxation 
times,\cite{speedy_hard_1998, stillinger_kauzmann_2001, 
aste_cell_2004, parisi_ideal_2005, 
angelani_configurational_2007, 
parisi_mean_field_2010, jadrich_equilibrium_2013}
while $I_{\text{vib}}^{\text{ind}}(N, \varphi, \varphi_J)$ increases rapidly 
with increasing $\varphi_J$ (if $\varphi$ is fixed as well). 
The last statement is just another formulation of the fact 
that basins of attraction 
decrease in volume when $\varphi$ approaches $\varphi_J$ for a fixed $N$.
Thus, the integrand in Eq. (\ref{eq:SaddlePointStart}) has a sharp maximum 
and we can replace the integral with 
$I_{\text{tot}}(N, \varphi) = N_J^{\text{ind}} (N, \varphi_{\text{DJ}}) 
I_{\text{vib}}^{\text{ind}}(N, \varphi, \varphi_{\text{DJ}}) w(N, \varphi)$,
where $\varphi_{\text{DJ}}$ is the ``dominant'' jamming density, 
given by the maximum of the integrand, 
and $w(N, \varphi)$ represents the ``width'' of the peak in the integrand.\cite{speedy_hard_1998, 
parisi_ideal_2005, parisi_mean_field_2010, berthier_microscopic_2011} 
It is usually believed that it is subexponential, 
so when we switch to entropies per particle (take the logarithm and divide by $N$),
the term with $w(N, \varphi)$ disappears.
Thus, we can just write 
\begin{equation}
    I_{\text{tot}}(N, \varphi) = N_J^{\text{ind}} (N, \varphi_{\text{DJ}}) 
    I_{\text{vib}}^{\text{ind}} (N, \varphi, \varphi_{\text{DJ}}).
  \label{eq:SaddlePoint}
\end{equation}
It means that we essentially have to discuss the same form of the separation 
into configurational and vibrational parts as in the 1D case: 
\begin{equation}
    S_{\text{tot}}(N, T, \varphi) = S_{\text{conf}} (N, \varphi_{\text{DJ}}) + S_{\text{vib}} (N, T, \varphi, \varphi_{\text{DJ}}).
  \label{eq:SeparationOfEntropies}
\end{equation}
We make some remarks on the function $\varphi_{\text{DJ}} (\varphi)$ in 
Appendix \ref{subsubsec:DominantJammingDensities}.


\subsection{\label{subsec:ThermodynamicIntegration} Vibrational entropy through thermodynamic integration}

Our aim here is to find whether $S_{\text{vib}} / k_B N$ contains the $s_{\text{mix}}$
and $\ln \left( \frac{V}{\Lambda^d N} \right)$ terms. 
Though we state as early as in the introduction that $s_{\text{mix}}$ 
shall not be present in $S_{\text{vib}} / k_B N$ due to compensating 
pre-integral terms, here we show
explicitly that $s_{\text{mix}}$ is not hidden in 
$S_{\text{vib}} / k_b N$ through the integral 
$\int_{\text{basin}} e^{-U_N(\vec{r})/k_B T} \text{d} \vec{r}$ either.
We choose a certain variant of thermodynamic 
integration for our purpose, 
but the possibility to practically implement 
it for real systems is of no concern to us.
It is a variant of a tether method of Speedy\cite{speedy_entropy_1993} or of a cell method of Donev \textit{et al.}\cite{donev_calculating_2007}

We want to find the volume of the available part of a certain basin of attraction (with the jamming density $\varphi_J$)
if the current volume fraction of the system is $\varphi$. At first, we do not multiply this volume by $\Pi_t N_t!$; 
thus, we want to find the vibrational partition function of a single basin 
$Z_{\text{vib}}^{\text{single}} = \frac{1}{\Pi_t N_t!} \frac{1}{\Lambda^{d N}} \int_{\text{basin}} e^{-U_N(\vec{r})/k_B T} \text{d} \vec{r} = 
\frac{1}{\Pi_t N_t!} \frac{1}{\Lambda^{d N}} I_{\text{vib}}^{\text{dist}}$ 
(while $Z_{\text{vib}} = \Pi_t N_t! Z_{\text{vib}}^{\text{single}}$).

We imagine that this basin of attraction in the PEL can somehow be ideally determined 
(\textit{e.g.}, by performing steepest descents in the PEL for all possible starting points\cite{ashwin_calculations_2012, asenjo_visualizing_2013}).
Then, we restrict the phase space to this particular basin. If a hard-sphere system during its dynamics 
reaches the boundary of this basin, it is elastically reflected from the boundary.

Now, we apply the tether method of Speedy.\cite{speedy_entropy_1993} We imagine that the center of each sphere is attached 
with a tether to a point where this center is located in the corresponding jammed configuration.
Alternatively, the centers of particles are surrounded with imaginary spherical cells, where the radius of a cell equals
the tether length for this sphere $L_i$.
When a sphere center reaches its cell wall during molecular dynamics, it is elastically reflected. 
For such a system, the vibrational Helmholtz free energy 
$A_{\text{vib}}^{\text{single}} = -k_B T \ln(Z_{\text{vib}}^{\text{single}})$ 
is additionally parameterized by radii of cells $L_i$. 
If we imply $L_i = \lambda R_i$, $A_{\text{vib}}^{\text{single}}$ is a function of $\lambda$.
$A_{\text{vib}}^{\text{single}}(\lambda = \infty)$ coincides with the vibrational free energy without cells.

Thermodynamic integration over $\lambda$ implies that the change in $A_{\text{vib}}^{\text{single}}(\lambda)$ is 
equal to the work that particle centers perform on the walls of their cells during the cell 
expansion,\cite{speedy_hard_1998, donev_calculating_2007} \textit{i.e.}
\begin{equation}
    A_{\text{vib}}^{\text{single}} = A_{\text{vib}}^{\text{single}}(\lambda_{\text{min}}) - 
    N \langle \int_{\lambda_{\text{min}}}^{\infty} p_c \text{d} \upsilon (\lambda) \rangle_{\text{cells}},
  \label{eq:IntegrationOverCells}
\end{equation}
where $p_c$ is the pressure on the cell walls and $\upsilon (\lambda)$ is the volume of a cell, 
$\upsilon_i (\lambda) = (4 / 3) \pi L_i^3 = \lambda^3 V_{\text{sp},i}$, where $V_{\text{sp},i}$ is the volume of the $i$th particle.

We can express the work on the wall of the cells through dimensionless quantities as 
$N k_B T \langle \int_{\lambda_{\text{min}}}^{\infty} Z_c \frac{\text{d} \upsilon (\lambda)}{\upsilon (\lambda)} \rangle_{\text{cells}}$,
where $Z_c = p_c \upsilon / k_B T$ is the reduced pressure on the cell walls. 
Reduced pressure is expressed in the general case as $Z = p V / N k_B T$,
but the pressure on each cell wall is counted from exactly one particle.

If $\lambda_{\text{min}}$ is sufficiently small and spheres located in minimal cells can never intersect, 
$Z_{\text{vib}}^{\text{single}} (\lambda_{\text{min}})$ can be expressed trivially as 
$\frac{1}{\Pi_t N_t!} \frac{1}{\Lambda^{d N}} \Pi_i^N (4 / 3) \pi L_i^3 = 
\frac{1}{\Pi_t N_t!} \frac{1}{\Lambda^{d N}} \Pi_i^N V_{\text{sp},i} \lambda_{\text{min}}^d = 
\frac{1}{\Pi_t N_t!} \frac{1}{\Lambda^{d N}} \lambda_{\text{min}}^{d N} \Pi_i^N V_{\text{sp},i}$.
The same result can be obtained from the fact that $Z_c = 1$ for $\lambda \in [0, \lambda_{\text{min}})$.

Now, if we switch to entropies per particle 
$S_{\text{vib}}^{\text{single}} / N k_B = U / N k_B T - A_{\text{vib}}^{\text{single}} / N k_B T$, 
we get (using Eq. (\ref{eq:PolydiperseStirling})) $S_{\text{vib}}^{\text{single}} / N k_B = 
\frac{U}{N k_B T} + 1 + s_{\text{mix}} - \ln(N) - \ln(\Lambda^d) + \ln(\lambda_{\text{min}}^d) + \langle \ln(V_{\text{sp}}) \rangle + 
\langle \int_{\lambda_{\text{min}}}^{\infty} Z_c \frac{\text{d} \upsilon (\lambda)}{\upsilon (\lambda)} \rangle$.

We would like to switch now to $\ln(\langle V_{\text{sp}} \rangle)$. We do this by simply writing
\begin{equation}
    \langle \ln(V_{\text{sp}}) \rangle = \alpha + \ln(\langle V_{\text{sp}} \rangle),
  \label{eq:SwitchingToAverageParticleVolume}
\end{equation}
where $\alpha = \langle \ln(V_{\text{sp}} / \langle V_{\text{sp}} \rangle) \rangle$ is 
some dimensionless quantity that characterizes the particle radii distribution.
Contrary to $s_{\text{mix}}$, it remains finite for all but very exotic distributions 
(\textit{cf}. Appendix \ref{subsec:RemarksOnVibrationalEntropy}).
For the monodisperse case, $\alpha = 0$.

After switching to $\langle V_{\text{sp}} \rangle$, we can introduce the density term $V/N$ given that 
$\varphi = N \langle V_{\text{sp}} \rangle / V$ and
$\ln(\langle V_{\text{sp}} \rangle) = \ln(V \varphi / N)$. We finally write
\begin{equation}
\begin{split}
    \frac{S_{\text{vib}}^{\text{single}}}{N k_B} =& 
    1 + s_{\text{mix}} - \ln(N) \\
    & + \frac{U}{N k_B T} + \ln \left( \frac{V}{\Lambda^d N} \right) 
    + \langle \ln \left( \frac{V_{\text{sp}}} {\langle V_{\text{sp}} \rangle} \right) \rangle \\
    & + \ln(\varphi \lambda_{\text{min}}^d) + 
    \langle \int_{\lambda_{\text{min}}}^{\infty} Z_c \frac{\text{d} \upsilon (\lambda)}{\upsilon (\lambda)} \rangle.
  \label{eq:SingleVibrationalEntropy}
\end{split}
\end{equation}

As mentioned, we are really interested in 
$S_{\text{vib}} / N k_B \equiv S_{\text{vib}}^{\text{ind}} / N k_B = S_{\text{vib}}^{\text{single}} / N k_B + \frac{1}{N}\ln(\Pi_t N_t!)$, where 
each basin of attraction is counted $\Pi_t N_t!$ times. The terms $1 + s_{\text{mix}} - \ln(N)$ in Eq. (\ref{eq:SingleVibrationalEntropy}) 
 cancel out exactly and we write
\begin{equation}
\begin{split}
    \frac{S_{\text{vib}}}{N k_B} =& 
    \frac{U}{N k_B T} + \ln \left( \frac{V}{\Lambda^d N} \right) + 
    \langle \ln \left( \frac{V_{\text{sp}}} {\langle V_{\text{sp}} \rangle} \right) \rangle \\
    & + \ln(\varphi \lambda_{\text{min}}^d) + 
    \langle \int_{\lambda_{\text{min}}}^{\infty} Z_c \frac{\text{d} \upsilon (\lambda)}{\upsilon (\lambda)} \rangle.
  \label{eq:VibrationalEntropy}
\end{split}
\end{equation}

Eq. (\ref{eq:VibrationalEntropy}) shows that $S_{\text{vib}} / N k_B$ contains $\ln \left( \frac{V}{\Lambda^d N} \right)$
and does not contain $s_{\text{mix}}$, which means that $s_{\text{mix}}$ shall be consumed by 
$S_{\text{conf}} / N k_B$ to make the relation $S_{\text{tot}} = S_{\text{conf}} + S_{\text{vib}}$ hold.
The choice of $\lambda_{\text{min}}$ is not particularly important: 
any changes in the choice of $\lambda_{\text{min}}$ will be incorporated by 
$\langle \int_{\lambda_{\text{min}}}^{\infty} Z_c \frac{\text{d} \upsilon (\lambda)}{\upsilon (\lambda)} \rangle$.
In Appendix \ref{subsec:RemarksOnVibrationalEntropy}, 
we demonstrate that the free volume equation of 
state\cite{kirkwood_critique_1950, buehler_free_1951, wood_note_1952} 
for hard spheres immediately follows from Eq. (\ref{eq:VibrationalEntropy}).
It is an approximate equation of state, but in the the limit 
$\varphi \to \varphi_J$ it asymptotically equals the 
``polytope'' equation of state by 
Salsburg and Wood,\cite{salsburg_equation_1962, stillinger_limiting_1969} 
which can be derived from first principles 
for the limit $\varphi \to \varphi_J$.
We also demonstrate in Appendix 
\ref{subsec:RemarksOnVibrationalEntropy} that $\alpha$ from 
Eq. (\ref{eq:SwitchingToAverageParticleVolume}) can not contain 
$s_{\text{mix}}$, even indirectly.

\subsection{Our resolution of the paradox}
\label{subsec:OurResolutionMain}

Eq. (\ref{eq:VibrationalEntropy}) shows that $S_{\text{vib}} / N k_B$ does not contain the $s_{\text{mix}}$ term 
but contains the $\ln \left( \frac{V}{\Lambda^d N} \right)$ term, which means that 
$s_{\text{mix}}$ from $S_{\text{tot}} / N k_B$ (Eq. (\ref{eq:TotalEntropy})) shall be consumed by 
$S_{\text{conf}} / N k_B$ to make the relation $S_{\text{tot}} = S_{\text{conf}} + S_{\text{vib}}$ hold.

The only remaining question is which entropy contains the unity from the ideal gas entropy per particle, Eq. (\ref{eq:IdealGasEntropy}).
This unity has to be removed from one of the terms ($S_{\text{vib}} / N k_B$ or $S_{\text{conf}} / N k_B$) if we 
subtract the ideal gas entropy $S_{\text{tot}}^{\circ}$ from $S_{\text{tot}}$.

The answer to this question is not provided by Eq. (\ref{eq:VibrationalEntropy}) directly, but we expect that 
this unity is contained in $S_{\text{vib}} / N k_B$, because that is what happens in the one-dimensional case.
Also, $s_{\text{mix}} = 0$ for the monodisperse case and we assume as usual that $S_{\text{conf}} (N, \varphi_J) / N k_B$ 
decreases with the increase in $\varphi_J$ and reaches zero at some 
$\varphi_J$.\cite{speedy_hard_1998, stillinger_kauzmann_2001, aste_cell_2004, parisi_ideal_2005, 
angelani_configurational_2007, parisi_mean_field_2010, jadrich_equilibrium_2013}  
If unity shall be subtracted from $S_{\text{conf}} (N, \varphi_J) / N k_B$, 
$S_{\text{conf}} (N, \varphi_J) / N k_B$ decreases to unity in the monodisperse case, not to zero, 
and this is physically unrealistic.

Thus, we suggest to write the relation $S_{\text{tot}} = S_{\text{conf}} + S_{\text{vib}}$ as 
\begin{equation}
    \Delta S_{\text{tot}} = \Delta_c S_{\text{conf}} + \Delta_v S_{\text{vib}},
    \label{eq:CorrectExcessEntropiesConclusion}
\end{equation}
where
\begin{equation}
\begin{aligned}
    &\frac{\Delta_c S_{\text{conf}}}{k_B N} = 
    \frac{S_{\text{conf}}}{k_B N} - s_{\text{mix}} \text{ and} \\
    &\frac{\Delta_v S_{\text{vib}}}{k_B N} = 
    \frac{S_{\text{vib}}}{k_B N} - 1 - \ln \left( \frac{V}{\Lambda^d N} \right) - \frac{U}{N k_B T}.
    \label{eq:CorrectExcessQuantitiesConclusion}
\end{aligned}
\end{equation}
All the quantities from Eq. (\ref{eq:CorrectExcessEntropiesConclusion}) 
if taken per particle are 
\begin{itemize}
\item 
{
    finite in the thermodynamic limit even for a continuous particle type distribution (polydispersity),
}
\item 
{
    continuous when introducing a small polydispersity to a monodisperse system,
}
\end{itemize}
Additionally, $\Delta_c S_{\text{conf}} (N, \varphi_J) / k_B N$ is supposed to decrease to zero with the increase of $\varphi_J$.
For hard spheres, all the terms from Eq. 
(\ref{eq:CorrectExcessEntropiesConclusion}) 
if taken per particle are also independent of the temperature $T$ and depend only on $\varphi$ and $\varphi_J$.

It may seem surprising that the terms from the ideal gas entropy $S_{\text{tot}}^{\circ}$ 
are distributed between $S_{\text{conf}}$ and $S_{\text{vib}}$, but exactly the same situation occurs for the well-known relation 
$\Delta S_{\text{tot}} / N k_B = \Delta Z - \langle \Delta \mu \rangle / k_B T$ 
(which stems from the expressions for the Gibbs free energy).\cite{carnahan_thermodynamic_1970, 
mansoori_equilibrium_1971, adams_chemical_1974, baranau_chemical_2016} Here, the unity from Eq. (\ref{eq:TotalEntropy})
is consumed in the excess reduced pressure $\Delta Z = Z - 1$, 
while $s_{\text{mix}} + \ln \left( \frac{V}{\Lambda^d N} \right)$ are consumed in the average excess chemical potential
$\langle \Delta \mu \rangle / k_B T$. Indeed, the ideal gas chemical potential for a single particle type 
$\mu_i^\circ$ is expressed\cite{baranau_chemical_2016} as 
$\frac{- \mu_i^\circ}{k_B T} = \ln \left( \frac{V}{\Lambda^d N_i} \right)$ and 
$\frac{- \langle \mu^\circ \rangle}{k_B T} = \sum_i \ln \left( \frac{V}{\Lambda^d N_i} \right) \frac{N_i}{N} = 
s_{\text{mix}} + \ln \left( \frac{V}{\Lambda^d N} \right)$. The $U / N k_B T$ term from Eq. (\ref{eq:TotalEntropy}) 
is actually consumed by the omitted $\Delta U / N k_B T$ term, which is always zero for hard spheres.

\section{The Adam--Gibbs and Random First Order Transition theories}
\label{sec:AdamGibbs}

The Adam--Gibbs (AG) relation\cite{adam_temperature_1965, 
cavagna_supercooled_2009, starr_relationship_2013} 
connects the relaxation time of a (glassy) isobaric 
system to the configurational entropy per particle:
\begin{equation}
    \tau_R = \tau_0 \exp\left( \frac{A}{T S_{\text{conf}} / N} \right),
    \label{eq:AdamGibbsOriginal}
\end{equation}
where $\tau_0$ and $A$ are constants. 
Eq. (\ref{eq:AdamGibbsOriginal}) corresponds to 
Eq. (21) in the original paper of Adam and Gibbs.\cite{adam_temperature_1965}
Eq. (21) in the original paper does not contain the number of particles, 
but $S_c$ in the original notation is the molar configurational entropy.
The relaxation time is often associated with the asymptotic alpha-relaxation 
time\cite{masri_dynamic_2009, brambilla_probing_2009, 
perez_angel_equilibration_2011, zaccarelli_polydispersity_2014} or the inverse diffusion 
coefficient.\cite{speedy_hard_1998, zaccarelli_polydispersity_2014}

As explained in Section \ref{subsec:Paradox}, $S_{\text{conf}} / N$ 
diverges (along with $s_{\text{mix}}$) 
in the thermodynamic limit for systems with 
continuous polydispersity. This alone indicates that the form 
(\ref{eq:AdamGibbsOriginal}) shall be amended (otherwise, $\tau_R \equiv 0$).
Additionally, $S_{\text{conf}} / N$ exhibits a jump when introducing a small polydispersity into a monodisperse system, and the corresponding jump 
would be induced into Eq. (\ref{eq:AdamGibbsOriginal}).
This is unphysical, because the relaxation dynamics shall remain almost unchanged
if a small polydispersity is introduced into a monodisperse system.
We demonstrated that $\Delta_c S_{\text{conf}} / N$ is always finite 
and continuous when introducing a small polydispersity 
into a monodisperse system. 
We thus believe that any relation connecting the relaxation time or 
an equivalent quantity to the configurational entropy per particle 
shall actually depend on $\Delta_c S_{\text{conf}} / N$:
\begin{equation}
    \tau_R = f(\Delta_c S_{\text{conf}} / N),
    \label{eq:AdamGibbsModifiedGeneral}
\end{equation}
Now, we provide a more elaborate explanation for the classical AG theory.

The AG theory assumes that a system is composed of relatively independent ``cooperatively rearranging regions'' (CRR), 
\textit{i.e.}, portions of the system that can undergo structural changes relatively independent of other regions, neighboring or not.
A structural change (or cooperative rearrangement) of a region means a transition between different states 
(different basins of attraction in the potential energy landscape of this region).
Some of the regions (subsystems) may not be able to perform structural changes because they have only one state available.
Adam and Gibbs assume that the relaxation rate of the entire system is proportional 
to the fraction of regions that can in principle undergo a structural change.
They arrive at the following equation (Eq. (11) in their original paper\cite{adam_temperature_1965}):
\begin{equation}
    \tau_R (T) = \tau_0 \exp \left( \frac{z^* \Delta \mu}{k_B T} \right),
    \label{eq:AdamGibbsRegionSize}
\end{equation}
where $\tau_0$ and $\Delta \mu$ are approximately independent of $T$, while $z^*$ 
is the minimal number of constituent elements in a subsystem that 
can undergo a structural change 
(constituent elements are molecules, 
monomeric segments in case of polymers, or hard spheres in case of colloids).

Then, Adam and Gibbs demonstrate that the configurational entropy of a cooperatively rearranging region $s_{\text{CRR}}$ is related 
to the number of its constituent elements as 
$z \frac{S_{\text{conf}}}{N} = s_{\text{CRR}}$, 
which is quite a natural result (to get $s_{\text{CRR}}$, 
we multiply the configurational entropy per particle 
by the number of particles in a region).
They write this relation directly for $z^*$:
\begin{equation}
    z^* \frac{S_{\text{conf}}}{N} = s_{\text{CRR}}^*,
    \label{eq:AdamGibbsRegionSizeAndEntropy}
\end{equation}
which is Eq. (20) in the original paper.\cite{adam_temperature_1965} 
$s_{\text{CRR}}^*$ here is the critical entropy 
corresponding to the minimum region size.
Note that the original notation is slightly different: 
the authors write the Avogadro number $N_A$ instead of $N$ 
(which is because they denote with $S_{\text{conf}}$
the molar configurational entropy), 
while $N$ in their paper denotes the number of cooperatively rearranging regions.

Next, Adam and Gibbs write the following: ``there must be a lower limit $z^*$ to the size of a 
cooperative subsystem that can perform a rearrangement into another configuration, 
this lower limit corresponding to a critical average number 
of configurations available to the subsystem. 
Certainly, this smallest size must be sufficiently large to have at least 
two configurations available to it
... For the following, however, we need not specify the 
numerical value of this small critical entropy $[s_{\text{CRR}}^*]$''.

We know now that for systems with continuous polydispersity both sides of 
Eq. (\ref{eq:AdamGibbsRegionSizeAndEntropy}) contain the diverging term 
$z^* k_B s_{\text{mix}}$. After canceling this diverging term on 
both sides we arrive at a relation where all the terms are 
well-behaving (finite and continuous):
\begin{equation}
    z^* = \frac{\Delta_c s_{\text{CRR}}^*} { \Delta_c S_{\text{conf}} / N}.
    \label{eq:AdamGibbsRegionSizeAndEntropyFixed}
\end{equation}
After substituting this result into Eq. (\ref{eq:AdamGibbsRegionSize}),
we obtain the modified AG relation
\begin{equation}
    \tau_R = \tau_0 \exp\left( \frac{A}{T \Delta_c S_{\text{conf}} / N} \right).
    \label{eq:AdamGibbsFixed}
\end{equation}
We note that $z^*$ can be too small to apply the Stirling approximation to 
$z^*!$ or at least to one of the constituent particle types 
to arrive at $s_{\text{mix}}$ as in Eq. (\ref{eq:PolydiperseStirling}). 
In this case one can imagine taking 
a large number of CRRs of size $z^*$, $N_\text{CRR}$, 
and writing Eq. (\ref{eq:AdamGibbsRegionSizeAndEntropy}) for all of them,
$N_\text{CRR} z^* \frac{S_{\text{conf}}}{N} = N_\text{CRR} s_{\text{CRR}}^*$.
For a sufficiently large $N_\text{CRR}$, the Stirling approximation applies
and we arrive at Eq. (\ref{eq:AdamGibbsRegionSizeAndEntropyFixed}) after 
canceling $N_\text{CRR}$ on both sides of the resulting equation.

The following example demonstrates why we have to subtract 
$k_B s_{\text{mix}}$ from $S_{\text{conf}} / N$ in the AG relation.
Suppose we have a monodisperse hard-sphere system ($s_{\text{mix}} = 0$) where 
minimum CRRs have a certain size $z^*$ corresponding to 
$s_{\text{CRR}}^* = k_B \ln(2)$. Thus, the number of local PEL minima 
of a minimum CRR (if all particles are treated as distinguishable) is 
$N_J^\text{dist} = 2 z^*!$.
Next, suppose that we introduce a small polydispersity to the system.
In the extreme case, we can just color the particles and postulate that 
we distinguish particles by color as well.
For a sufficiently small polydispersity (or for coloring), 
$N_J^\text{dist}$ shall remain unchanged,
$N_J^\text{dist} = 2 z^*!$, 
because the structure of basins remains almost unchanged.
On the contrary, $s_{\text{CRR}}^*$ shall be counted as if particles with equal 
radii (or color) are treated as indistinguishable. Thus, 
$s_{\text{CRR}}^* = k_B \frac{2 z^*!}{\Pi_t z^*_t !}$, where $z^*_t$
is the number of particles of type $t$ among $z^*$.
After applying the Stirling approximation to the nominator
and Eq. (\ref{eq:PolydiperseStirling}) to the denominator, we obtain
$s_{\text{CRR}}^* = k_B \ln(2) + k_B z^* s_{\text{mix}}$.
After canceling $k_B z^* s_{\text{mix}}$ on both sides of 
Eq. (\ref{eq:AdamGibbsRegionSizeAndEntropy}), we once again 
obtain Eq. (\ref{eq:AdamGibbsRegionSizeAndEntropyFixed}).
If $z^*$ is too small to apply the Stirling approximation to the nominator 
or denominator, we can imagine analyzing many CRRs simultaneously
and then cancel $N_{\text{CRR}}$,
as suggested in the previous paragraph.

Now, we briefly justify the usage of $\Delta_c S_{\text{conf}} / N$
in the relaxation time prediction 
from the Random First Order Transition (mosaic) 
theory.\cite{kirkpatrick_random_2015, cavagna_supercooled_2009, 
starr_relationship_2013, bouchaud_adam_2004}
The equation for the relaxation time from this 
theory looks similar to the original AG relation (\ref{eq:AdamGibbsOriginal}): 
\begin{equation}
\tau_R = \tau_0 \exp \left( C 
\frac
{ Y(T)^{\frac{d}{d - \theta}} }
{ T \left[ T S_{\text{conf}} / V \right]^{\frac{\theta}{d - \theta}} } 
\right), 
\label{eq:RfotOriginal}
\end{equation}
where 
$Y(T)$ is the generalized surface tension coefficient, $\theta$ is the 
parameter of the theory, and 
$\frac{S_{\text{conf}}}{V} = \frac{S_{\text{conf}}}{N} \frac{N}{V}$.
Due to the presence of $\frac{S_{\text{conf}}}{N}$, 
Eq. (\ref{eq:RfotOriginal}) possesses the same 
problems as Eq. (\ref{eq:AdamGibbsOriginal}): 
$\tau_R \equiv 0$ for systems 
with continuous polydispersity and $\tau_R$ is 
discontinuous when introducing a small 
polydispersity to a monodisperse system.
Similarly to the AG relation, we suggest that $S_{\text{conf}} / N$
shall be replaced by $\Delta_c S_{\text{conf}} / N$.
Indeed, the theory assumes that a (glassy) system consists of 
a patchwork (mosaic) of different metastable regions, 
while transitions between different system states occur via nucleation 
of such metastable regions (entropic droplets).
Their growth is hindered by the 
surface tension with neighboring regions, 
and the free energy loss at a droplet radius $R$ due to this tension is 
$\Delta F_{\text{loss}} \sim Y R^\theta$. 
For the usual surface tension, $\theta = d - 1$, 
but the theory only implies that $\theta \leq d - 1$.
At the same time, it is postulated that droplets get the ``entropic'' 
free energy gain $\Delta F_{\text{gain}} \sim 
-T \frac{S_{\text{conf}} }{V} R^d$: when a droplet transitions from an ``unstable'' state with many available configurations 
into a metastable state with a single 
(on an experimental timescale) available 
configuration (up to permutations), 
the free energy corresponding to the configurational entropy 
of the droplet is released.
Thus, the droplet final radius $R_0$ is obtained from 
$\Delta F_{\text{loss}} = \Delta F_{\text{gain}}$ and the 
free energy barrier of nucleation $\Delta$ equals the maximum value of 
$\Delta F_{\text{loss}} + \Delta F_{\text{gain}}$ for $R \leq R_0$, 
which leads to $\Delta \sim \frac
{ Y(T)^{\frac{d}{d - \theta}} }
{ \left[ T S_{\text{conf}} / V \right]^{\frac{\theta}{d - \theta}} }$,
which after substitution into $\tau_R = \tau_0 \exp(\Delta / k_B T)$
produces Eq. (\ref{eq:RfotOriginal}).
As already noted, $S_{\text{conf}} / N$ or $S_{\text{conf}} / V$ are 
poorly-behaving quantities, divergent and discontinuous. 
Hence, we suggest that the entropic gain shall be calculated 
through the well-behaving quantity $\Delta_c S_{\text{conf}} / V$.
Indeed, the $k_B s_{\text{mix}}$ part of 
$S_{\text{conf}} / N$ (if calculated per particle) is 
always present in any part of the system 
just due to system 
composition and can not be released as the free energy during 
the growth of metastable entropic droplets 
(and in any other process if a system remains uniform in composition). 
The ``entropic'' free energy gain can thus happen 
only up to $k_B s_{\text{mix}}$ (per particle) 
and shall in fact be expressed as 
$\Delta F_{\text{gain}} \sim -T \frac{\Delta_c S_{\text{conf}} }{V} R^d$.
Eq. (\ref{eq:RfotOriginal}) shall thus be written as
\begin{equation}
\tau_R = \tau_0 \exp \left( C 
\frac
{ Y(T)^{\frac{d}{d - \theta}} }
{ T \left[ T \Delta_c S_{\text{conf}} / V \right]^{\frac{\theta}{d - \theta}} } 
\right), 
\label{eq:RfotFixed}
\end{equation}

Even if one follows the resolution of Ozawa and Berthier and redefines 
the configurational entropy by essentially redefining the mixing entropy, 
the redefined mixing entropy is still inaccessible to the entropic free energy gain 
during the growth of metastable droplets. Thus, the redefined mixing entropy 
shall still be subtracted from the configurational entropy 
in Eq. (\ref{eq:RfotFixed}) (as well as in Eq. (\ref{eq:AdamGibbsFixed})).

Finally, we specify how Eq. (\ref{eq:AdamGibbsRegionSizeAndEntropyFixed}) 
shall look for a hard-sphere system
(following Ref.\cite{speedy_hard_1998} but accounting for $\Delta_c$).
For a system of hard spheres, 
we can express the reduced pressure as 
$Z = p V / N k_B T = p \langle V_{\text{sp}} \rangle / k_B T \varphi$,
where $\langle V_{\text{sp}} \rangle$ is the average sphere volume. 
The isobaric assumption of the AG theory ($p = \text{const}$) 
implies that in Eq. (\ref{eq:AdamGibbsRegionSizeAndEntropyFixed}) 
$A / T = C \varphi Z(\varphi)$, 
where $C = A p / k_B V_{\text{sp}} = \text{const}$. 
We consequently write for hard spheres 
\begin{equation}
    \tau_R(\varphi) = \tau_0 
    \exp\left( C \frac{\varphi Z(\varphi)}{\Delta_c S_{\text{conf}} (\varphi) / N} \right),
  \label{eq:AdamGibbsFixedForHardSpheres}
\end{equation}
where $S_{\text{conf}} (\varphi) = S_{\text{conf}} (\varphi_{\text{DJ}} (\varphi))$ is the equilibrium complexity (\textit{cf}. Eq. (\ref{eq:SeparationOfEntropies})).

\section{Conclusions}

In this paper, we suggest that a natural way to write the relation $S_{\text{tot}} = S_{\text{conf}} + S_{\text{vib}}$ is
\begin{equation}
    \Delta S_{\text{tot}} = \Delta_c S_{\text{conf}} + \Delta_v S_{\text{vib}},
    \label{eq:CorrectExcessEntropiesConclusion}
\end{equation}
where
\begin{equation}
\begin{aligned}
    &\frac{\Delta_c S_{\text{conf}}}{k_B N} = 
    \frac{S_{\text{conf}}}{k_B N} - s_{\text{mix}} \text{ and} \\
    &\frac{\Delta_v S_{\text{vib}}}{k_B N} = 
    \frac{S_{\text{vib}}}{k_B N} - 1 - \ln \left( \frac{V}{\Lambda^d N} \right) - \frac{U}{N k_B T}.
    \label{eq:CorrectExcessQuantitiesConclusion}
\end{aligned}
\end{equation}
All the quantities from Eq. (\ref{eq:CorrectExcessEntropiesConclusion}) 
if taken per particle are 
\begin{itemize}
\item 
{
    finite in the thermodynamic limit even for a continuous particle type distribution (polydispersity),
}
\item 
{
    continuous when introducing a small polydispersity to a monodisperse system,
}
\end{itemize}
Additionally, $\Delta_c S_{\text{conf}} (N, \varphi_J) / k_B N$ is supposed to decrease to zero with the increase of $\varphi_J$.

This resolution does not require any redefinition of basins of attraction and 
is in line with usual treatment of $S_{\text{tot}}$, when only $\Delta S_{\text{tot}}$ is discussed instead.
One may argue that this is merely a technical rewriting of the equation for entropies, but we think that working with some 
sorts of delta-quantities lies in the nature of entropy. 
Information entropy for an arbitrary distribution 
(basically, $s_{\text{mix}}$) diverges when switching to continuous distributions. 
Thus, information entropy for 
continuous distributions is represented in the information theory 
through the differential entropy, 
which for an arbitrary function is its information entropy 
with respect to the uniform distribution in a unit interval $[0, 1]$.\cite{stone_information_2015, lazo_entropy_1978}
In statistical physics, we can use even a more natural approach: 
measure entropies with respect to the ideal gas of a corresponding particle size distribution. 
This paper essentially discusses how exactly we have to 
distribute the terms from the ideal gas entropy $S_{\text{tot}}^{\circ}$ between $S_{\text{conf}}$ and $S_{\text{vib}}$.

We also demonstrated that the Adam--Gibbs and the Random First Order Transition 
theory relations for the relaxation time of (glassy) systems 
shall be written through $\Delta_c S_{\text{conf}} / N$ 
or $\Delta_c S_{\text{conf}} / V$ instead of $S_{\text{conf}} / N$ or 
$S_{\text{conf}} / V$, respectively:
\begin{equation}
\begin{aligned}
    \tau_R^{\text{AG}} =& 
    \tau_0 \exp\left( \frac{A}{T \Delta_c S_{\text{conf}} / N} \right),\\
    \tau_R^{\text{RFOT}} =& 
    \tau_0 \exp \left( C 
    \frac
    { Y(T)^{\frac{d}{d - \theta}} }
    { T \left[ T \Delta_c S_{\text{conf}} / V \right]^{\frac{\theta}{d - \theta}} } 
    \right).
\end{aligned}
\end{equation}
In general, we suggest that any relation that expresses the relaxation time through 
$S_{\text{conf}}$ shall in fact depend on $\Delta_c S_{\text{conf}} / N$:
\begin{equation}
    \tau_R = f(\Delta_c S_{\text{conf}} / N).
\end{equation}

Our final remark is on how to interpret previous papers 
that rely on the separation of entropies.
If a paper writes out the expression for entropies as 
$\Delta S_{\text{tot}} = S_{\text{conf}} + \Delta S_{\text{vib}}$ or implies it, 
one has to read this relation rather as 
$\Delta S_{\text{tot}} = \Delta_c S_{\text{conf}} + \Delta_v S_{\text{vib}}$.
If the authors used the relation $S_{\text{conf}} / N = 0$ to define 
the density of the ideal glass transition or of the glass close packing limit, 
this relation just has to be reinterpreted as 
$\Delta_c S_{\text{conf}} / N = 0$ and the estimated location of 
either the ideal glass transition or the glass close packing limit 
shall be kept unchanged, 
though special care shall be taken of course on 
how exactly the calculations were performed.
Similarly, if the authors used the Adam--Gibbs or 
Random First Order Transition (mosaic) theories 
for validating the values of $S_{\text{conf}} / N$ against 
measured relaxation times or for fitting 
some unknown parameters, it can well be that these results hold, 
but one has to read $\Delta_c S_{\text{conf}}$ instead of $S_{\text{conf}}$
everywhere in the paper, including the AG or mosaic relations.

The presented results can be useful in understanding the Edwards entropy\cite{edwards_theory_1989, bowles_edwards_2011, 
asenjo_numerical_2014, bi_statistical_2015, baranau_upper_2016} for polydisperse systems in granular matter studies.
For frictionless particles, the Edwards entropy is equivalent under some definitions to 
the configurational entropy.

\section*{Acknowledgments}
We thank Misaki Ozawa and Ludovic Berthier for helpful and insightful discussions as well as comments on the manuscript. 
We thank Patrick Charbonneau for reading the manuscript.
We thank Sibylle N\"agle for preparing Fig. \ref{fig:1D_PhaseSpace}.
We are also grateful to the two anonymous reviewers 
of the manuscript for their comments and suggestions.

\appendix
\section*{Appendix: Remarks on the vibrational entropy and statistical physics of glasses and hard spheres}

\subsection{Remarks on the vibrational entropy: free volume theory and polytopes}
\label{subsec:RemarksOnVibrationalEntropy}

One may ask whether the $\alpha = \langle \ln(V_{\text{sp}} / \langle V_{\text{sp}} \rangle) \rangle$
term from Eq. (\ref{eq:SwitchingToAverageParticleVolume}) 
somehow contains $s_{\text{mix}}$ indirectly. 
It does not, because it remains finite 
for all but very exotic continuous distributions, 
contrary to $s_{\text{mix}}$.
Indeed, when discretizing a particle radii distribution 
$f(r)$ with a step $\delta$,
$\alpha = \int f(r) \ln \left( \frac{r^d}{\langle r^d \rangle} \right) \text{d} r$ when $\delta \to 0$, 
which is fundamentally different from $s_{\text{mix}}$ in Eq. (\ref{eq:MixingEntropyDivergence}), 
which contains the diverging $\ln(\delta)$ term. Additionally, $\alpha$ is continuous when introducing a small polydispersity to 
a monodisperse system, contrary to $s_{\text{mix}}$. In general, $s_{\text{mix}}$ can be made 
arbitrary different from $\alpha$---for example, by introducing particle types with radii infinitely close 
to some existing particle types. Then, $\alpha$ will remain almost unchanged, while $s_{\text{mix}}$ can be changed arbitrarily.
As an extreme example, one can introduce particle types by coloring colloidal particles and postulating that 
we distinguish particles by color as well as by radii. Then, $\alpha$ will remain exactly the same, while $s_{\text{mix}}$
will change. This example shows the difference in the nature of $s_{\text{mix}}$ and $\alpha$: $s_{\text{mix}}$ stems from our conventions on 
indistinguishability and $\alpha$---from geometrical radii.

We can easily determine the largest possible $\lambda_{\text{min}}$ in Eq. (\ref{eq:VibrationalEntropy}). 
We can take a jammed configuration at $\varphi_J$, 
then scale particle radii linearly by a factor $(\varphi / \varphi_J)^{1/d}$ to ensure the density $\varphi$. 
Now, to maintain the original particle radii, we scale the entire system (particle radii and distances between particles) 
by $(\varphi_J / \varphi)^{1/d}$. The density of such a system is still $\varphi$, 
the shrunk particles possess the original particle radii $R_i$, 
and the original particles are enlarged and possess the radii $(\varphi_J / \varphi)^{1/d} R_i$.
These enlarged particles can be treated as initial cells for the tether/cell method. Such cells will be ``jammed'', 
because the original particles were jammed at $\varphi_J$. The lengths of tethers are then 
$(\varphi_J / \varphi)^{1/d} R_i - R_i$.
Thus, a natural choice for $\lambda_{\text{min}} = (\varphi_J / \varphi)^{1/d} - 1$ and 
the $\ln(\varphi \lambda_{\text{min}}^d)$ term from Eqs. (\ref{eq:SingleVibrationalEntropy}) and (\ref{eq:VibrationalEntropy})
looks like
\begin{equation}
    \ln(\varphi \lambda_{\text{min}}^d) = d \ln(\varphi_J^{1/d} - \varphi^{1/d}).
  \label{eq:FreeVolumeLambda}
\end{equation}

If $\varphi$ approaches $\varphi_J$, particles are hardly able to move further 
away from tether centers than prescribed by $\lambda_{\text{min}}$ from Eq. (\ref{eq:FreeVolumeLambda}).
It means that we can assume $Z_c \approx 0$ for $\lambda > \lambda_{\text{min}}$ and thus write
\begin{equation}
    \langle \int_{\lambda_{\text{min}}}^{\infty} Z_c \frac{\text{d} \upsilon (\lambda)}{\upsilon (\lambda)} \rangle = 0.
  \label{eq:FreeVolumeZeroIntegral}
\end{equation}
in Eqs. (\ref{eq:SingleVibrationalEntropy}) and (\ref{eq:VibrationalEntropy}), making these equations completely analytical.
Eqs. (\ref{eq:SingleVibrationalEntropy}) and (\ref{eq:VibrationalEntropy}) will then essentially represent 
the free volume theory\cite{kirkwood_critique_1950, buehler_free_1951, wood_note_1952} for the polydisperse case, because Eq. (\ref{eq:FreeVolumeLambda}) 
essentially describes such free volumes. Eqs. (\ref{eq:SingleVibrationalEntropy}), (\ref{eq:VibrationalEntropy}), 
and (\ref{eq:FreeVolumeLambda}) show that $S_{\text{vib}} / N k_B$ (or $S_{\text{vib}}^{\text{single}} / N k_B$) 
behave with $\varphi$ in exactly the same way as for the monodisperse case in the free volume approximation 
(up to the size distribution-dependent constant $\alpha = \langle \ln \left( \frac{V_{\text{sp}}} {\langle V_{\text{sp}} \rangle} \right) \rangle$).

In the same way as we write $p = - (\partial A_{\text{tot}} / \partial V)_{N,T} = 
k_B T (\partial \ln Z_{\text{tot}} / \partial V)_{N,T}$ for the total free energy and the entire phase space, 
we can introduce the glass pressure $p_g$ if we assume that only a particular basin of attraction is left in the phase space 
(\textit{cf}. Appendix \ref{subsubsec:GlassPressure}).
Glass pressure in this formulation has been studied, among other works, 
in the papers of Speedy\cite{speedy_hard_1998} and Donev, Stillinger, and Torquato.\cite{donev_configurational_2007}
We write $p_g = k_B T (\partial \ln Z_{\text{vib}} / \partial V)_{N,T}$.
Repeating the steps from Appendix \ref{subsubsec:TotalEntropyThroughPressures}, we write in the same way
$Z_g (\varphi, \varphi_J) = p_g V / N k_B T = 1 - \varphi \frac{\partial \Delta S_{\text{vib}} / N k_B}{\partial \varphi}$, 
where $\Delta S_{\text{vib}} = S_{\text{vib}} - S_{\text{tot}}^{\circ}$. One can also use $S_{\text{vib}}^{\text{single}}$, 
depending on the context---$Z_g$ does not depend on this choice.
Using Eqs. (\ref{eq:VibrationalEntropy}), (\ref{eq:FreeVolumeLambda}), 
and (\ref{eq:IdealGasEntropy}), we obtain for the polydisperse case $Z_g = 1 + \frac{1}{(\varphi_J / \varphi)^{1/d} - 1}$, 
which is a well-known free volume glass equation of state 
(previously derived for the monodisperse case, though).\cite{buehler_free_1951, wood_note_1952, salsburg_equation_1962}

When Speedy\cite{speedy_entropy_1993} and Donev \textit{et al.}\cite{donev_calculating_2007} 
applied the original tether/cell methods, they could not ideally determine basins of attraction 
(the tether method would actually be quite useless in that case).
Still, it is natural to assume that up to a certain $\lambda$ the system is not be able to (quickly) leave the original basin of attraction.
Thus, these authors performed the integration in Eqs. (\ref{eq:SingleVibrationalEntropy}) or (\ref{eq:VibrationalEntropy}) 
up to a certain $\lambda_{\text{max}}$. $\lambda_{\text{max}}$ was determined 
by a jump in the measured cell pressure.\cite{speedy_hard_1998, donev_configurational_2007} 
Such a jump indicates that the system starts to explore other basins of attraction.
Some more advanced corrections, like extrapolating $Z_c(\lambda)$, can also be utilized.

Finally, we note that it is known that the basin of attraction approaches a 
polytope when $\varphi \to \varphi_J$.\cite{salsburg_equation_1962, torquato_jammed_2010}
For the monodisperse case, a glass equation of state has been derived long ago\cite{salsburg_equation_1962} for polytopes and slightly later a 
complete form of the polytope free energy $A_{\text{vib}}$ was obtained.\cite{stillinger_limiting_1969}
The polytope glass equation of state
is equivalent to the free volume one for $\varphi \to \varphi_J$ and looks like $Z_g = 1 + \frac{d}{(\varphi_J / \varphi) - 1}$.\cite{salsburg_equation_1962}  
We found that it was easier for our purposes to amend the
tether/cell method to the polydisperse case than to amend the complete computation of $A_{\text{vib}}$ through polytope geometries. 

\subsection{Remarks on statistical physics of glasses and hard spheres}
\label{subsec:AppendixStatisticalPhysics}

In this section, we use entropies per particle 
$s_{\text{tot}} = S_{\text{tot}} / k_B N$, 
$s_{\text{vib}} = S_{\text{vib}} / k_B N$, and
$s_{\text{conf}} = S_{\text{conf}} / k_B N$.
It is convenient, because 
$\Delta s_{\text{tot}} = s_{\text{tot}} - s_{\text{tot}}^\circ$ 
is truly a function of $\varphi$ only,
$\Delta_v s_{\text{vib}} = s_{\text{vib}} - 1 - \ln \left( \frac{V}{\Lambda^d N} \right) - \frac{U}{N k_B T}$
is truly a function of $\varphi$ and $\varphi_J$ only 
(as well as $\Delta s_{\text{vib}} = s_{\text{vib}} - s_{\text{tot}}^\circ$), 
and $s_{\text{conf}}$ and 
$\Delta_c s_{\text{conf}} = s_{\text{conf}} - s_{\text{mix}}$
are truly functions of $\varphi_J$ only.

\subsubsection{Total entropy through pressures}
\label{subsubsec:TotalEntropyThroughPressures}

Equilibrium fluid pressure $p$ is routinely defined in the canonical ensemble 
through the Helmholtz free energy $A_{\text{tot}} = -k_B T \ln Z_{\text{tot}}$ as 
$p = - (\partial A_{\text{tot}} / \partial V)_{N,T}$. This relation essentially defines pressures through 
the partition function: $p =  k_B T (\partial \ln Z_{\text{tot}} / \partial V)_{N,T}$. 
We use the reduced pressure (compressibility factor) $Z = p V / N k_B T$.
For hard spheres, it is possible to express $Z$ through the excess entropy per particle $\Delta s_{\text{tot}}$. 
Specifically, by using the relations $A = U - T S_{\text{tot}}$ and $U = \frac{3}{2} N k_B T$, 
we write $p = T (\partial S_{\text{tot}} / \partial V)_{N,T}$.
After utilizing $S_{\text{tot}} = \Delta S_{\text{tot}} + S_{\text{tot}}^\circ$, we get $p = T (\partial \Delta S_{\text{tot}} / \partial V)_{N,T} + 
T (\partial S_{\text{tot}}^\circ / \partial V)_{N,T}$. The last term is the ideal gas pressure $N k_B T$.
If we switch to the reduced pressure $Z = p V / N k_B T$, 
we obtain $Z = V \left( \frac{\partial \Delta s_{\text{tot}} }{\partial V} \right)_{N,T} + 1$. 
By replacing $V$ with $\varphi$ through $\varphi = N \langle V_{\text{sp}} \rangle / V$ 
(where $\langle V_{\text{sp}} \rangle$ is the average sphere volume), we finally write: 
\begin{equation}
    Z (\varphi) = 1 - \varphi \frac{\text{d} \Delta s_{\text{tot}} (\varphi) }{\text{d} \varphi}.
  \label{eq:PressureThroughEntropy}
\end{equation}
Integration of Eq. (\ref{eq:PressureThroughEntropy}) leads to 
$\Delta s_{\text{tot}}(\varphi) = \Delta s_{\text{tot}}(\varphi_0) - 
\int \limits_{\varphi_0}^\varphi \frac{Z(\varphi') - 1}{\varphi'} \text{d} \varphi'$.
If we use the ideal gas as the reference state, we get
\begin{equation}
    \Delta s_{\text{tot}}(\varphi) = - \int \limits_{0}^\varphi \frac{Z(\varphi') - 1}{\varphi'} \text{d} \varphi'.
  \label{eq:EntropyThroughPressure}
\end{equation}

\subsubsection{Glass pressure and vibrational entropy}
\label{subsubsec:GlassPressure}

In this subsection, we study the relationships between the glass pressure and the vibrational entropy.
Glass pressure in the present formulation has been studied, among other works, 
in the papers of Speedy\cite{speedy_hard_1998} and Donev, Stillinger, and Torquato,\cite{donev_configurational_2007}
but without the corrections in the vibrational entropy needed in the polydisperse case.
In the same way as we write $p = - (\partial A_{\text{tot}} / \partial V)_{N,T} = 
k_B T (\partial \ln Z_{\text{tot}} / \partial V)_{N,T}$ for the total free energy and the entire phase space, 
we can introduce glass pressure $p_g$ if we assume that only a particular basin of attraction 
is left in the phase space.\cite{speedy_hard_1998, donev_configurational_2007}
We write $p_g = k_B T (\partial \ln Z_{\text{vib}} / \partial V)_{N,T}$.
Repeating the steps from Appendix \ref{subsubsec:TotalEntropyThroughPressures}, we write in the same way
$Z_g (\varphi, \varphi_J) = p_g V / N k_B T$ and 
\begin{equation}
\begin{aligned}
    Z_g (\varphi, \varphi_J) =& 
    1 - \varphi \frac{\partial \Delta s_{\text{vib}} (\varphi, \varphi_J)}{\partial \varphi} \\
    =& 1 - \varphi \frac{\partial \Delta_v s_{\text{vib}} (\varphi, \varphi_J)}{\partial \varphi},
  \label{eq:GlassyPressureThroughVibrationalEntropy}
\end{aligned}
\end{equation}
One can also use $s_{\text{vib}}^{\text{single}}$, 
depending on the context---$Z_g$ does not depend on this choice.

If one wants to measure $Z_g$, this definition assumes that one has to track during the system evolution (molecular dynamics) 
time points when the system crosses the boundary of a basin of attraction and to elastically reflect the velocity hypervector from this boundary. 
For example, one can perform a steepest descent in the pseudo-PEL at each particle collision 
during the event-driven molecular dynamics simulation. If the basin is changed between collisions, 
one has to find with the binary search the time between the last collisions when the basin is switched from one to another.
This procedure is computationally expensive, but presumably tractable for small systems.
The basin of attraction does not have to dominate the phase space at a given density to define and measure $Z_g$, 
what is required is that a system can be equilibrated inside this particular basin, if only this basin is left in the phase space.
It is usually assumed that non-ergodicity at high densities stems from hindered movement of a system between basins,
not inside basins, so we assume that it is always possible to equilibrate a system inside a single basin.

Note that one can use either $\Delta$ or $\Delta_v$ in Eq. (\ref{eq:GlassyPressureThroughVibrationalEntropy}), it produces the same $Z_g$ 
(while $\frac{\partial s_{\text{mix}} }{\partial \varphi} = 0$).
The purpose of using $\Delta$ or $\Delta_v$ in Eq. 
(\ref{eq:GlassyPressureThroughVibrationalEntropy}) as well as in 
Eq. (\ref{eq:PressureThroughEntropy}) is to remove the 
$\ln \left( \frac{V}{\Lambda^d N} \right)$ term to be able to use 
derivatives over $\varphi$ correctly.
Eq. (\ref{eq:EntropyThroughPressure}) uses the ideal gas state once again 
(along the $\Delta$ usage)---as a starting configuration for the 
thermodynamic integration, 
but these are two ``independent'' usages. Similarly, writing 
$\Delta s_{\text{vib}}$ in Eq. (\ref{eq:GlassyPressureThroughVibrationalEntropy}) 
does not mean that we use the ideal gas as a starting point for the 
thermodynamic integration---it just means that we measure $s_{\text{vib}}$ 
with respect to the ideal gas.
Indeed, integration of Eq. (\ref{eq:GlassyPressureThroughVibrationalEntropy}) shall 
rather start from the jammed configuration and produces 
\begin{equation}
    \Delta_v s_{\text{vib}}(\varphi, \varphi_J) = \Delta_v s_0(\varphi_J) + 
    \int \limits_{\varphi}^{\varphi_J} \frac{Z_g(\varphi', \varphi_J) - 1}{\varphi'} \text{d} \varphi'.
  \label{eq:VibrationalEntropyThroughPressureAppendix}
\end{equation}
By comparing Eqs. (\ref{eq:VibrationalEntropyThroughPressureAppendix}) and 
(\ref{eq:VibrationalEntropy}) we conclude that 
$\Delta_v s_0(\varphi_J) = \langle \ln \left( \frac{V_{\text{sp}}} {\langle V_{\text{sp}} \rangle} \right) \rangle + A$, where $A$ represents the geometry of the 
polytope.
If we use $Z_g$ from the polytope theory, we get 
$\Delta_v s_{\text{vib}}(\varphi, \varphi_J) = \Delta_v s_0(\varphi_J) + 
d \ln(\varphi_J - \varphi)$, 
slightly different from the free volume 
theory, $\Delta_v s_{\text{vib}}(\varphi, \varphi_J) = \Delta_v s_0(\varphi_J) + 
d \ln(\varphi_J^{1/d} - \varphi^{1/d})$, where 
$\Delta_v s_0(\varphi_J) = \langle \ln \left( \frac{V_{\text{sp}}} {\langle V_{\text{sp}} \rangle} \right) \rangle + \langle \int_{\lambda_{\text{min}}}^{\infty} Z_c \frac{\text{d} \upsilon (\lambda)}{\upsilon (\lambda)} \rangle$ 
(Eqs. (\ref{eq:VibrationalEntropy}) and (\ref{eq:FreeVolumeLambda})).
Note that Speedy wrote\cite{speedy_hard_1998} the equation for ``polytope'' vibrational entropies as 
$\Delta s_{\text{vib}}(\varphi, \varphi_J) = \Delta s_0(\varphi_J) + 
d \ln(\varphi_J - \varphi)$, which is technically correct but 
conceals the point that $s_{\text{mix}}$ is extra-removed from $s_{\text{vib}}(\varphi, \varphi_J)$.

\subsubsection{Dominant jamming densities}
\label{subsubsec:DominantJammingDensities}

Here, we make some remarks on the jamming density that dominates the phase space 
at a given $\varphi$, the dominant jamming density $\varphi_{\text{DJ}} (\varphi)$ 
from Eq. (\ref{eq:SeparationOfEntropies}).
Its value is determined by the maximum of the integrand in 
Eq. (\ref{eq:SaddlePointStart}) and thus by
\begin{equation}
    \left[ \frac{\text{d} \Delta_c s_{\text{conf}} (\varphi_J) }{\text{d} \varphi_J} + 
    \frac{\partial \Delta_v s_{\text{vib}} (\varphi, \varphi_J) }{\partial \varphi_J} \right]_{\varphi_J = \varphi_{\text{DJ}}} = 0.
  \label{eq:SaddlePointMaximumAppendix}
\end{equation}

It is useful to investigate the behavior of the ``dominant glass reduced pressure'' 
$Z_g(\varphi, \varphi_{\text{DJ}}(\varphi))$. 
It was done by Donev, Stillinger, and Torquato,\cite{donev_configurational_2007} but the necessity to 
work with $\Delta_v s_{\text{vib}}$ instead of $\Delta s_{\text{vib}}$ in the polydisperse case 
was not realized at that time, so we repeat their derivation with the corresponding changes.

According to Eq. (\ref{eq:GlassyPressureThroughVibrationalEntropy}),
we need to investigate the behavior of 
$\frac{\partial \Delta_v s_{\text{vib}} (\varphi, \varphi_{\text{DJ}})}{\partial \varphi}$.
To do this, we subtract $S_{\text{tot}}^\circ$ from 
Eq. (\ref{eq:SeparationOfEntropies}), divide it by $k_B N$, 
and fully differentiate it:
$\frac{\text{d} \Delta s_{\text{tot}} (\varphi)} {\text{d} \varphi} = 
\left[ \frac{\text{d} \Delta_c s_{\text{conf}}(\varphi_J)} {\text{d} \varphi_J} \right]_{\varphi_J = \varphi_{\text{DJ}}}
    \frac{\text{d} \varphi_{\text{DJ}}}{\text{d} \varphi}
+ \frac{\partial \Delta_v s_{\text{vib}} (\varphi, \varphi_{\text{DJ}})} {\partial \varphi} + 
\left[ \frac{\partial \Delta_v s_{\text{vib}} (\varphi, \varphi_J)} {\partial \varphi_J} \right]_{\varphi_J = \varphi_{\text{DJ}}}
\frac{\text{d} \varphi_{\text{DJ}}}{\text{d} \varphi} = 
\frac{\partial \Delta_v s_{\text{vib}}(\varphi, \varphi_{\text{DJ}})} {\partial \varphi} + 
\frac{\text{d} \varphi_{\text{DJ}}}{\text{d} \varphi} 
\left[ \frac{\text{d} \Delta_c s_{\text{conf}}(\varphi_J)} {\text{d} \varphi_J} +
\frac{\partial \Delta_v s_{\text{vib}}(\varphi, \varphi_J)} {\partial \varphi_J} \right]_{\varphi_J = \varphi_{\text{DJ}}}$.
According to Eq. (\ref{eq:SaddlePointMaximumAppendix}), the term in square brackets shall be zero and thus
$\frac{\text{d} \Delta s_{\text{tot}} (\varphi)} {\text{d} \varphi} = \frac{\partial \Delta_v s_{\text{vib}}(\varphi, \varphi_{\text{DJ}})} {\partial \varphi}$.
By comparing this result with Eqs. (\ref{eq:PressureThroughEntropy}) and 
(\ref{eq:GlassyPressureThroughVibrationalEntropy}), we conclude that as long as the 
phase space is ergodic (the saddle point approximation holds, $\varphi$ is below the ideal glass transition density)
\begin{equation}
    Z(\varphi) = Z_g(\varphi, \varphi_{\text{DJ}}(\varphi)).
  \label{eq:EqualityOfPressuresAppendix}
\end{equation}
In other words, the equilibrium (fluid) reduced pressure equals the glass reduced pressure of the dominant 
basins of attraction.

\bibliography{ConfigurationalEntropyParadox}

\begin{thebibliography}{60}%
\makeatletter
\providecommand \@ifxundefined [1]{%
 \@ifx{#1\undefined}
}%
\providecommand \@ifnum [1]{%
 \ifnum #1\expandafter \@firstoftwo
 \else \expandafter \@secondoftwo
 \fi
}%
\providecommand \@ifx [1]{%
 \ifx #1\expandafter \@firstoftwo
 \else \expandafter \@secondoftwo
 \fi
}%
\providecommand \natexlab [1]{#1}%
\providecommand \enquote  [1]{``#1''}%
\providecommand \bibnamefont  [1]{#1}%
\providecommand \bibfnamefont [1]{#1}%
\providecommand \citenamefont [1]{#1}%
\providecommand \href@noop [0]{\@secondoftwo}%
\providecommand \href [0]{\begingroup \@sanitize@url \@href}%
\providecommand \@href[1]{\@@startlink{#1}\@@href}%
\providecommand \@@href[1]{\endgroup#1\@@endlink}%
\providecommand \@sanitize@url [0]{\catcode `\\12\catcode `\$12\catcode
  `\&12\catcode `\#12\catcode `\^12\catcode `\_12\catcode `\%12\relax}%
\providecommand \@@startlink[1]{}%
\providecommand \@@endlink[0]{}%
\providecommand \url  [0]{\begingroup\@sanitize@url \@url }%
\providecommand \@url [1]{\endgroup\@href {#1}{\urlprefix }}%
\providecommand \urlprefix  [0]{URL }%
\providecommand \Eprint [0]{\href }%
\providecommand \doibase [0]{http://dx.doi.org/}%
\providecommand \selectlanguage [0]{\@gobble}%
\providecommand \bibinfo  [0]{\@secondoftwo}%
\providecommand \bibfield  [0]{\@secondoftwo}%
\providecommand \translation [1]{[#1]}%
\providecommand \BibitemOpen [0]{}%
\providecommand \bibitemStop [0]{}%
\providecommand \bibitemNoStop [0]{.\EOS\space}%
\providecommand \EOS [0]{\spacefactor3000\relax}%
\providecommand \BibitemShut  [1]{\csname bibitem#1\endcsname}%
\let\auto@bib@innerbib\@empty
\bibitem [{\citenamefont {Speedy}(1998{\natexlab{a}})}]{speedy_hard_1998}%
  \BibitemOpen
  \bibfield  {author} {\bibinfo {author} {\bibfnamefont {R.~J.}\ \bibnamefont
  {Speedy}},\ }\href {\doibase 10.1080/00268979809483148} {\bibfield  {journal}
  {\bibinfo  {journal} {Mol. Phys.}\ }\textbf {\bibinfo {volume} {95}},\
  \bibinfo {pages} {169} (\bibinfo {year} {1998}{\natexlab{a}})}\BibitemShut
  {NoStop}%
\bibitem [{\citenamefont {Stillinger}, \citenamefont {Debenedetti},\ and\
  \citenamefont {Truskett}(2001)}]{stillinger_kauzmann_2001}%
  \BibitemOpen
  \bibfield  {author} {\bibinfo {author} {\bibfnamefont {F.~H.}\ \bibnamefont
  {Stillinger}}, \bibinfo {author} {\bibfnamefont {P.~G.}\ \bibnamefont
  {Debenedetti}}, \ and\ \bibinfo {author} {\bibfnamefont {T.~M.}\ \bibnamefont
  {Truskett}},\ }\href {\doibase 10.1021/jp011840i} {\bibfield  {journal}
  {\bibinfo  {journal} {J. Phys. Chem. B}\ }\textbf {\bibinfo {volume} {105}},\
  \bibinfo {pages} {11809} (\bibinfo {year} {2001})}\BibitemShut {NoStop}%
\bibitem [{\citenamefont {Sastry}(2001)}]{sastry_relationship_2001}%
  \BibitemOpen
  \bibfield  {author} {\bibinfo {author} {\bibfnamefont {S.}~\bibnamefont
  {Sastry}},\ }\href {\doibase 10.1038/35051524} {\bibfield  {journal}
  {\bibinfo  {journal} {Nature}\ }\textbf {\bibinfo {volume} {409}},\ \bibinfo
  {pages} {164} (\bibinfo {year} {2001})}\BibitemShut {NoStop}%
\bibitem [{\citenamefont {Angelani}\ and\ \citenamefont
  {Foffi}(2007)}]{angelani_configurational_2007}%
  \BibitemOpen
  \bibfield  {author} {\bibinfo {author} {\bibfnamefont {L.}~\bibnamefont
  {Angelani}}\ and\ \bibinfo {author} {\bibfnamefont {G.}~\bibnamefont
  {Foffi}},\ }\href {\doibase 10.1088/0953-8984/19/25/256207} {\bibfield
  {journal} {\bibinfo  {journal} {J. Phys.: Condens. Matter}\ }\textbf
  {\bibinfo {volume} {19}},\ \bibinfo {pages} {256207} (\bibinfo {year}
  {2007})}\BibitemShut {NoStop}%
\bibitem [{\citenamefont {Donev}, \citenamefont {Stillinger},\ and\
  \citenamefont {Torquato}(2007{\natexlab{a}})}]{donev_configurational_2007}%
  \BibitemOpen
  \bibfield  {author} {\bibinfo {author} {\bibfnamefont {A.}~\bibnamefont
  {Donev}}, \bibinfo {author} {\bibfnamefont {F.~H.}\ \bibnamefont
  {Stillinger}}, \ and\ \bibinfo {author} {\bibfnamefont {S.}~\bibnamefont
  {Torquato}},\ }\href {\doibase doi:10.1063/1.2775928} {\bibfield  {journal}
  {\bibinfo  {journal} {J. Chem. Phys.}\ }\textbf {\bibinfo {volume} {127}},\
  \bibinfo {pages} {124509} (\bibinfo {year} {2007}{\natexlab{a}})}\BibitemShut
  {NoStop}%
\bibitem [{\citenamefont {Foffi}\ and\ \citenamefont
  {Angelani}(2008)}]{foffi_short_2008}%
  \BibitemOpen
  \bibfield  {author} {\bibinfo {author} {\bibfnamefont {G.}~\bibnamefont
  {Foffi}}\ and\ \bibinfo {author} {\bibfnamefont {L.}~\bibnamefont
  {Angelani}},\ }\href {\doibase 10.1088/0953-8984/20/7/075108} {\bibfield
  {journal} {\bibinfo  {journal} {J. Phys.: Condens. Matter}\ }\textbf
  {\bibinfo {volume} {20}},\ \bibinfo {pages} {075108} (\bibinfo {year}
  {2008})}\BibitemShut {NoStop}%
\bibitem [{\citenamefont {Starr}, \citenamefont {Douglas},\ and\ \citenamefont
  {Sastry}(2013)}]{starr_relationship_2013}%
  \BibitemOpen
  \bibfield  {author} {\bibinfo {author} {\bibfnamefont {F.~W.}\ \bibnamefont
  {Starr}}, \bibinfo {author} {\bibfnamefont {J.~F.}\ \bibnamefont {Douglas}},
  \ and\ \bibinfo {author} {\bibfnamefont {S.}~\bibnamefont {Sastry}},\ }\href
  {\doibase 10.1063/1.4790138} {\bibfield  {journal} {\bibinfo  {journal} {J.
  Chem. Phys.}\ }\textbf {\bibinfo {volume} {138}},\ \bibinfo {pages} {12A541}
  (\bibinfo {year} {2013})}\BibitemShut {NoStop}%
\bibitem [{\citenamefont {Asenjo}, \citenamefont {Paillusson},\ and\
  \citenamefont {Frenkel}(2014)}]{asenjo_numerical_2014}%
  \BibitemOpen
  \bibfield  {author} {\bibinfo {author} {\bibfnamefont {D.}~\bibnamefont
  {Asenjo}}, \bibinfo {author} {\bibfnamefont {F.}~\bibnamefont {Paillusson}},
  \ and\ \bibinfo {author} {\bibfnamefont {D.}~\bibnamefont {Frenkel}},\ }\href
  {\doibase 10.1103/PhysRevLett.112.098002} {\bibfield  {journal} {\bibinfo
  {journal} {Phys. Rev. Lett.}\ }\textbf {\bibinfo {volume} {112}},\ \bibinfo
  {pages} {098002} (\bibinfo {year} {2014})}\BibitemShut {NoStop}%
\bibitem [{\citenamefont {Martiniani}\ \emph {et~al.}(2016)\citenamefont
  {Martiniani}, \citenamefont {Schrenk}, \citenamefont {Stevenson},
  \citenamefont {Wales},\ and\ \citenamefont
  {Frenkel}}]{martiniani_turning_2016}%
  \BibitemOpen
  \bibfield  {author} {\bibinfo {author} {\bibfnamefont {S.}~\bibnamefont
  {Martiniani}}, \bibinfo {author} {\bibfnamefont {K.~J.}\ \bibnamefont
  {Schrenk}}, \bibinfo {author} {\bibfnamefont {J.~D.}\ \bibnamefont
  {Stevenson}}, \bibinfo {author} {\bibfnamefont {D.~J.}\ \bibnamefont
  {Wales}}, \ and\ \bibinfo {author} {\bibfnamefont {D.}~\bibnamefont
  {Frenkel}},\ }\href {\doibase 10.1103/PhysRevE.93.012906} {\bibfield
  {journal} {\bibinfo  {journal} {Phys. Rev. E}\ }\textbf {\bibinfo {volume}
  {93}},\ \bibinfo {pages} {012906} (\bibinfo {year} {2016})}\BibitemShut
  {NoStop}%
\bibitem [{\citenamefont {Frenkel}(2014)}]{frenkel_why_2014}%
  \BibitemOpen
  \bibfield  {author} {\bibinfo {author} {\bibfnamefont {D.}~\bibnamefont
  {Frenkel}},\ }\href {\doibase 10.1080/00268976.2014.904051} {\bibfield
  {journal} {\bibinfo  {journal} {Mol. Phys.}\ }\textbf {\bibinfo {volume}
  {112}},\ \bibinfo {pages} {2325} (\bibinfo {year} {2014})}\BibitemShut
  {NoStop}%
\bibitem [{\citenamefont {Meng}\ \emph {et~al.}(2010)\citenamefont {Meng},
  \citenamefont {Arkus}, \citenamefont {Brenner},\ and\ \citenamefont
  {Manoharan}}]{meng_free_2010}%
  \BibitemOpen
  \bibfield  {author} {\bibinfo {author} {\bibfnamefont {G.}~\bibnamefont
  {Meng}}, \bibinfo {author} {\bibfnamefont {N.}~\bibnamefont {Arkus}},
  \bibinfo {author} {\bibfnamefont {M.~P.}\ \bibnamefont {Brenner}}, \ and\
  \bibinfo {author} {\bibfnamefont {V.~N.}\ \bibnamefont {Manoharan}},\ }\href
  {\doibase 10.1126/science.1181263} {\bibfield  {journal} {\bibinfo  {journal}
  {Science}\ }\textbf {\bibinfo {volume} {327}},\ \bibinfo {pages} {560}
  (\bibinfo {year} {2010})}\BibitemShut {NoStop}%
\bibitem [{\citenamefont {Ozawa}\ and\ \citenamefont
  {Berthier}(2017)}]{ozawa_does_2017}%
  \BibitemOpen
  \bibfield  {author} {\bibinfo {author} {\bibfnamefont {M.}~\bibnamefont
  {Ozawa}}\ and\ \bibinfo {author} {\bibfnamefont {L.}~\bibnamefont
  {Berthier}},\ }\href {\doibase 10.1063/1.4972525} {\bibfield  {journal}
  {\bibinfo  {journal} {J. Chem. Phys.}\ }\textbf {\bibinfo {volume} {146}},\
  \bibinfo {pages} {014502} (\bibinfo {year} {2017})}\BibitemShut {NoStop}%
\bibitem [{\citenamefont {Baranau}\ and\ \citenamefont
  {Tallarek}(2016)}]{baranau_chemical_2016}%
  \BibitemOpen
  \bibfield  {author} {\bibinfo {author} {\bibfnamefont {V.}~\bibnamefont
  {Baranau}}\ and\ \bibinfo {author} {\bibfnamefont {U.}~\bibnamefont
  {Tallarek}},\ }\href {\doibase 10.1063/1.4953079} {\bibfield  {journal}
  {\bibinfo  {journal} {J. Chem. Phys.}\ }\textbf {\bibinfo {volume} {144}},\
  \bibinfo {pages} {214503} (\bibinfo {year} {2016})}\BibitemShut {NoStop}%
\bibitem [{\citenamefont {Stone}(2015)}]{stone_information_2015}%
  \BibitemOpen
  \bibfield  {author} {\bibinfo {author} {\bibfnamefont {J.~V.}\ \bibnamefont
  {Stone}},\ }\href@noop {} {\emph {\bibinfo {title} {Information Theory: A
  Tutorial Introduction}}},\ \bibinfo {edition} {1st}\ ed.\ (\bibinfo
  {publisher} {Sebtel Press},\ \bibinfo {address} {England},\ \bibinfo {year}
  {2015})\BibitemShut {NoStop}%
\bibitem [{\citenamefont {Lazo}\ and\ \citenamefont
  {Rathie}(1978)}]{lazo_entropy_1978}%
  \BibitemOpen
  \bibfield  {author} {\bibinfo {author} {\bibfnamefont {A.~V.}\ \bibnamefont
  {Lazo}}\ and\ \bibinfo {author} {\bibfnamefont {P.}~\bibnamefont {Rathie}},\
  }\href {\doibase 10.1109/TIT.1978.1055832} {\bibfield  {journal} {\bibinfo
  {journal} {IEEE Transact. Inform. Theory}\ }\textbf {\bibinfo {volume}
  {24}},\ \bibinfo {pages} {120} (\bibinfo {year} {1978})}\BibitemShut
  {NoStop}%
\bibitem [{\citenamefont {Stillinger}\ and\ \citenamefont
  {Salsburg}(1969)}]{stillinger_limiting_1969}%
  \BibitemOpen
  \bibfield  {author} {\bibinfo {author} {\bibfnamefont {F.~H.}\ \bibnamefont
  {Stillinger}}\ and\ \bibinfo {author} {\bibfnamefont {Z.~W.}\ \bibnamefont
  {Salsburg}},\ }\href {\doibase 10.1007/BF01007250} {\bibfield  {journal}
  {\bibinfo  {journal} {J. Stat. Phys.}\ }\textbf {\bibinfo {volume} {1}},\
  \bibinfo {pages} {179} (\bibinfo {year} {1969})}\BibitemShut {NoStop}%
\bibitem [{\citenamefont {Speedy}(1993)}]{speedy_entropy_1993}%
  \BibitemOpen
  \bibfield  {author} {\bibinfo {author} {\bibfnamefont {R.~J.}\ \bibnamefont
  {Speedy}},\ }\href {\doibase 10.1080/00268979300102911} {\bibfield  {journal}
  {\bibinfo  {journal} {Mol. Phys.}\ }\textbf {\bibinfo {volume} {80}},\
  \bibinfo {pages} {1105} (\bibinfo {year} {1993})}\BibitemShut {NoStop}%
\bibitem [{\citenamefont {Speedy}(1998{\natexlab{b}})}]{speedy_pressure_1998}%
  \BibitemOpen
  \bibfield  {author} {\bibinfo {author} {\bibfnamefont {R.~J.}\ \bibnamefont
  {Speedy}},\ }\href {\doibase 10.1088/0953-8984/10/20/006} {\bibfield
  {journal} {\bibinfo  {journal} {J. Phys.: Condens. Matter}\ }\textbf
  {\bibinfo {volume} {10}},\ \bibinfo {pages} {4387} (\bibinfo {year}
  {1998}{\natexlab{b}})}\BibitemShut {NoStop}%
\bibitem [{\citenamefont {Charbonneau}\ \emph {et~al.}(2014)\citenamefont
  {Charbonneau}, \citenamefont {Kurchan}, \citenamefont {Parisi}, \citenamefont
  {Urbani},\ and\ \citenamefont {Zamponi}}]{charbonneau_fractal_2014}%
  \BibitemOpen
  \bibfield  {author} {\bibinfo {author} {\bibfnamefont {P.}~\bibnamefont
  {Charbonneau}}, \bibinfo {author} {\bibfnamefont {J.}~\bibnamefont
  {Kurchan}}, \bibinfo {author} {\bibfnamefont {G.}~\bibnamefont {Parisi}},
  \bibinfo {author} {\bibfnamefont {P.}~\bibnamefont {Urbani}}, \ and\ \bibinfo
  {author} {\bibfnamefont {F.}~\bibnamefont {Zamponi}},\ }\href {\doibase
  10.1038/ncomms4725} {\bibfield  {journal} {\bibinfo  {journal} {Nat.
  Commun.}\ }\textbf {\bibinfo {volume} {5}},\ \bibinfo {pages} {3725}
  (\bibinfo {year} {2014})}\BibitemShut {NoStop}%
\bibitem [{\citenamefont {Charbonneau}\ \emph {et~al.}(2017)\citenamefont
  {Charbonneau}, \citenamefont {Kurchan}, \citenamefont {Parisi}, \citenamefont
  {Urbani},\ and\ \citenamefont {Zamponi}}]{charbonneau_glass_2017}%
  \BibitemOpen
  \bibfield  {author} {\bibinfo {author} {\bibfnamefont {P.}~\bibnamefont
  {Charbonneau}}, \bibinfo {author} {\bibfnamefont {J.}~\bibnamefont
  {Kurchan}}, \bibinfo {author} {\bibfnamefont {G.}~\bibnamefont {Parisi}},
  \bibinfo {author} {\bibfnamefont {P.}~\bibnamefont {Urbani}}, \ and\ \bibinfo
  {author} {\bibfnamefont {F.}~\bibnamefont {Zamponi}},\ }\href {\doibase
  10.1146/annurev-conmatphys-031016-025334} {\bibfield  {journal} {\bibinfo
  {journal} {Annu. Rev. Condens. Matter Phys.}\ }\textbf {\bibinfo {volume}
  {8}},\ \bibinfo {pages} {265} (\bibinfo {year} {2017})}\BibitemShut {NoStop}%
\bibitem [{\citenamefont {Stillinger}, \citenamefont {DiMarzio},\ and\
  \citenamefont {Kornegay}(1964)}]{stillinger_systematic_1964}%
  \BibitemOpen
  \bibfield  {author} {\bibinfo {author} {\bibfnamefont {F.~H.}\ \bibnamefont
  {Stillinger}}, \bibinfo {author} {\bibfnamefont {E.~A.}\ \bibnamefont
  {DiMarzio}}, \ and\ \bibinfo {author} {\bibfnamefont {R.~L.}\ \bibnamefont
  {Kornegay}},\ }\href {\doibase 10.1063/1.1725362} {\bibfield  {journal}
  {\bibinfo  {journal} {J. Chem. Phys.}\ }\textbf {\bibinfo {volume} {40}},\
  \bibinfo {pages} {1564} (\bibinfo {year} {1964})}\BibitemShut {NoStop}%
\bibitem [{\citenamefont {Stillinger}(1995)}]{stillinger_topographic_1995}%
  \BibitemOpen
  \bibfield  {author} {\bibinfo {author} {\bibfnamefont {F.~H.}\ \bibnamefont
  {Stillinger}},\ }\href {\doibase 10.1126/science.267.5206.1935} {\bibfield
  {journal} {\bibinfo  {journal} {Science}\ }\textbf {\bibinfo {volume}
  {267}},\ \bibinfo {pages} {1935} (\bibinfo {year} {1995})}\BibitemShut
  {NoStop}%
\bibitem [{\citenamefont {Debenedetti}\ and\ \citenamefont
  {Stillinger}(2001)}]{debenedetti_supercooled_2001}%
  \BibitemOpen
  \bibfield  {author} {\bibinfo {author} {\bibfnamefont {P.~G.}\ \bibnamefont
  {Debenedetti}}\ and\ \bibinfo {author} {\bibfnamefont {F.~H.}\ \bibnamefont
  {Stillinger}},\ }\href {\doibase 10.1038/35065704} {\bibfield  {journal}
  {\bibinfo  {journal} {Nature}\ }\textbf {\bibinfo {volume} {410}},\ \bibinfo
  {pages} {259} (\bibinfo {year} {2001})}\BibitemShut {NoStop}%
\bibitem [{\citenamefont {Torquato}\ and\ \citenamefont
  {Jiao}(2010)}]{torquato_robust_2010}%
  \BibitemOpen
  \bibfield  {author} {\bibinfo {author} {\bibfnamefont {S.}~\bibnamefont
  {Torquato}}\ and\ \bibinfo {author} {\bibfnamefont {Y.}~\bibnamefont
  {Jiao}},\ }\href {\doibase 10.1103/PhysRevE.82.061302} {\bibfield  {journal}
  {\bibinfo  {journal} {Phys. Rev. E}\ }\textbf {\bibinfo {volume} {82}},\
  \bibinfo {pages} {061302} (\bibinfo {year} {2010})}\BibitemShut {NoStop}%
\bibitem [{\citenamefont {Ashwin}\ \emph {et~al.}(2012)\citenamefont {Ashwin},
  \citenamefont {Blawzdziewicz}, \citenamefont {O'Hern},\ and\ \citenamefont
  {Shattuck}}]{ashwin_calculations_2012}%
  \BibitemOpen
  \bibfield  {author} {\bibinfo {author} {\bibfnamefont {S.~S.}\ \bibnamefont
  {Ashwin}}, \bibinfo {author} {\bibfnamefont {J.}~\bibnamefont
  {Blawzdziewicz}}, \bibinfo {author} {\bibfnamefont {C.~S.}\ \bibnamefont
  {O'Hern}}, \ and\ \bibinfo {author} {\bibfnamefont {M.~D.}\ \bibnamefont
  {Shattuck}},\ }\href {\doibase 10.1103/PhysRevE.85.061307} {\bibfield
  {journal} {\bibinfo  {journal} {Phys. Rev. E}\ }\textbf {\bibinfo {volume}
  {85}},\ \bibinfo {pages} {061307} (\bibinfo {year} {2012})}\BibitemShut
  {NoStop}%
\bibitem [{\citenamefont {Baranau}\ and\ \citenamefont
  {Tallarek}(2014{\natexlab{a}})}]{baranau_random_close_2014}%
  \BibitemOpen
  \bibfield  {author} {\bibinfo {author} {\bibfnamefont {V.}~\bibnamefont
  {Baranau}}\ and\ \bibinfo {author} {\bibfnamefont {U.}~\bibnamefont
  {Tallarek}},\ }\href {\doibase 10.1039/C3SM52959B} {\bibfield  {journal}
  {\bibinfo  {journal} {Soft Matter}\ }\textbf {\bibinfo {volume} {10}},\
  \bibinfo {pages} {3826} (\bibinfo {year} {2014}{\natexlab{a}})}\BibitemShut
  {NoStop}%
\bibitem [{\citenamefont {Zinchenko}(1994)}]{zinchenko_algorithm_1994}%
  \BibitemOpen
  \bibfield  {author} {\bibinfo {author} {\bibfnamefont {A.}~\bibnamefont
  {Zinchenko}},\ }\href {\doibase 10.1006/jcph.1994.1168} {\bibfield  {journal}
  {\bibinfo  {journal} {J. Comput. Phys.}\ }\textbf {\bibinfo {volume} {114}},\
  \bibinfo {pages} {298} (\bibinfo {year} {1994})}\BibitemShut {NoStop}%
\bibitem [{\citenamefont {Asenjo}\ \emph {et~al.}(2013)\citenamefont {Asenjo},
  \citenamefont {Stevenson}, \citenamefont {Wales},\ and\ \citenamefont
  {Frenkel}}]{asenjo_visualizing_2013}%
  \BibitemOpen
  \bibfield  {author} {\bibinfo {author} {\bibfnamefont {D.}~\bibnamefont
  {Asenjo}}, \bibinfo {author} {\bibfnamefont {J.~D.}\ \bibnamefont
  {Stevenson}}, \bibinfo {author} {\bibfnamefont {D.~J.}\ \bibnamefont
  {Wales}}, \ and\ \bibinfo {author} {\bibfnamefont {D.}~\bibnamefont
  {Frenkel}},\ }\href {\doibase 10.1021/jp312457a} {\bibfield  {journal}
  {\bibinfo  {journal} {J. Phys. Chem. B}\ }\textbf {\bibinfo {volume} {117}},\
  \bibinfo {pages} {12717} (\bibinfo {year} {2013})}\BibitemShut {NoStop}%
\bibitem [{\citenamefont {Carnahan}\ and\ \citenamefont
  {Starling}(1970)}]{carnahan_thermodynamic_1970}%
  \BibitemOpen
  \bibfield  {author} {\bibinfo {author} {\bibfnamefont {N.~F.}\ \bibnamefont
  {Carnahan}}\ and\ \bibinfo {author} {\bibfnamefont {K.~E.}\ \bibnamefont
  {Starling}},\ }\href {\doibase 10.1063/1.1674033} {\bibfield  {journal}
  {\bibinfo  {journal} {J. Chem. Phys.}\ }\textbf {\bibinfo {volume} {53}},\
  \bibinfo {pages} {600} (\bibinfo {year} {1970})}\BibitemShut {NoStop}%
\bibitem [{\citenamefont {Mansoori}\ \emph {et~al.}(1971)\citenamefont
  {Mansoori}, \citenamefont {Carnahan}, \citenamefont {Starling},\ and\
  \citenamefont {Leland}}]{mansoori_equilibrium_1971}%
  \BibitemOpen
  \bibfield  {author} {\bibinfo {author} {\bibfnamefont {G.~A.}\ \bibnamefont
  {Mansoori}}, \bibinfo {author} {\bibfnamefont {N.~F.}\ \bibnamefont
  {Carnahan}}, \bibinfo {author} {\bibfnamefont {K.~E.}\ \bibnamefont
  {Starling}}, \ and\ \bibinfo {author} {\bibfnamefont {T.~W.}\ \bibnamefont
  {Leland}},\ }\href {\doibase doi:10.1063/1.1675048} {\bibfield  {journal}
  {\bibinfo  {journal} {J. Chem. Phys.}\ }\textbf {\bibinfo {volume} {54}},\
  \bibinfo {pages} {1523} (\bibinfo {year} {1971})}\BibitemShut {NoStop}%
\bibitem [{\citenamefont {Adams}(1974)}]{adams_chemical_1974}%
  \BibitemOpen
  \bibfield  {author} {\bibinfo {author} {\bibfnamefont {D.~J.}\ \bibnamefont
  {Adams}},\ }\href {\doibase 10.1080/00268977400102551} {\bibfield  {journal}
  {\bibinfo  {journal} {Mol. Phys.}\ }\textbf {\bibinfo {volume} {28}},\
  \bibinfo {pages} {1241} (\bibinfo {year} {1974})}\BibitemShut {NoStop}%
\bibitem [{\citenamefont {Donev}, \citenamefont {Stillinger},\ and\
  \citenamefont {Torquato}(2007{\natexlab{b}})}]{donev_calculating_2007}%
  \BibitemOpen
  \bibfield  {author} {\bibinfo {author} {\bibfnamefont {A.}~\bibnamefont
  {Donev}}, \bibinfo {author} {\bibfnamefont {F.~H.}\ \bibnamefont
  {Stillinger}}, \ and\ \bibinfo {author} {\bibfnamefont {S.}~\bibnamefont
  {Torquato}},\ }\href {\doibase 10.1016/j.jcp.2006.12.013} {\bibfield
  {journal} {\bibinfo  {journal} {J. Comput. Phys.}\ }\textbf {\bibinfo
  {volume} {225}},\ \bibinfo {pages} {509} (\bibinfo {year}
  {2007}{\natexlab{b}})}\BibitemShut {NoStop}%
\bibitem [{\citenamefont {Adam}\ and\ \citenamefont
  {Gibbs}(1965)}]{adam_temperature_1965}%
  \BibitemOpen
  \bibfield  {author} {\bibinfo {author} {\bibfnamefont {G.}~\bibnamefont
  {Adam}}\ and\ \bibinfo {author} {\bibfnamefont {J.~H.}\ \bibnamefont
  {Gibbs}},\ }\href {\doibase 10.1063/1.1696442} {\bibfield  {journal}
  {\bibinfo  {journal} {J. Chem. Phys.}\ }\textbf {\bibinfo {volume} {43}},\
  \bibinfo {pages} {139} (\bibinfo {year} {1965})}\BibitemShut {NoStop}%
\bibitem [{\citenamefont {Cavagna}(2009)}]{cavagna_supercooled_2009}%
  \BibitemOpen
  \bibfield  {author} {\bibinfo {author} {\bibfnamefont {A.}~\bibnamefont
  {Cavagna}},\ }\href {\doibase 10.1016/j.physrep.2009.03.003} {\bibfield
  {journal} {\bibinfo  {journal} {Phys. Rep.}\ }\textbf {\bibinfo {volume}
  {476}},\ \bibinfo {pages} {51} (\bibinfo {year} {2009})}\BibitemShut
  {NoStop}%
\bibitem [{\citenamefont {Kirkpatrick}\ and\ \citenamefont
  {Thirumalai}(2015)}]{kirkpatrick_random_2015}%
  \BibitemOpen
  \bibfield  {author} {\bibinfo {author} {\bibfnamefont {T.}~\bibnamefont
  {Kirkpatrick}}\ and\ \bibinfo {author} {\bibfnamefont {D.}~\bibnamefont
  {Thirumalai}},\ }\href {\doibase 10.1103/RevModPhys.87.183} {\bibfield
  {journal} {\bibinfo  {journal} {Rev. Mod. Phys.}\ }\textbf {\bibinfo {volume}
  {87}},\ \bibinfo {pages} {183} (\bibinfo {year} {2015})}\BibitemShut
  {NoStop}%
\bibitem [{\citenamefont {Bouchaud}\ and\ \citenamefont
  {Biroli}(2004)}]{bouchaud_adam_2004}%
  \BibitemOpen
  \bibfield  {author} {\bibinfo {author} {\bibfnamefont {J.-P.}\ \bibnamefont
  {Bouchaud}}\ and\ \bibinfo {author} {\bibfnamefont {G.}~\bibnamefont
  {Biroli}},\ }\href {\doibase 10.1063/1.1796231} {\bibfield  {journal}
  {\bibinfo  {journal} {J. Chem. Phys.}\ }\textbf {\bibinfo {volume} {121}},\
  \bibinfo {pages} {7347} (\bibinfo {year} {2004})}\BibitemShut {NoStop}%
\bibitem [{\citenamefont {Baranau}\ \emph {et~al.}(2016)\citenamefont
  {Baranau}, \citenamefont {Zhao}, \citenamefont {Scheel}, \citenamefont
  {Tallarek},\ and\ \citenamefont {Schr\"oter}}]{baranau_upper_2016}%
  \BibitemOpen
  \bibfield  {author} {\bibinfo {author} {\bibfnamefont {V.}~\bibnamefont
  {Baranau}}, \bibinfo {author} {\bibfnamefont {S.-C.}\ \bibnamefont {Zhao}},
  \bibinfo {author} {\bibfnamefont {M.}~\bibnamefont {Scheel}}, \bibinfo
  {author} {\bibfnamefont {U.}~\bibnamefont {Tallarek}}, \ and\ \bibinfo
  {author} {\bibfnamefont {M.}~\bibnamefont {Schr\"oter}},\ }\href {\doibase
  10.1039/C6SM00567E} {\bibfield  {journal} {\bibinfo  {journal} {Soft Matter}\
  }\textbf {\bibinfo {volume} {12}},\ \bibinfo {pages} {3991} (\bibinfo {year}
  {2016})}\BibitemShut {NoStop}%
\bibitem [{\citenamefont {Frenkel}\ and\ \citenamefont
  {Ladd}(1984)}]{frenkel_new_1984}%
  \BibitemOpen
  \bibfield  {author} {\bibinfo {author} {\bibfnamefont {D.}~\bibnamefont
  {Frenkel}}\ and\ \bibinfo {author} {\bibfnamefont {A.~J.~C.}\ \bibnamefont
  {Ladd}},\ }\href {\doibase 10.1063/1.448024} {\bibfield  {journal} {\bibinfo
  {journal} {J. Chem. Phys.}\ }\textbf {\bibinfo {volume} {81}},\ \bibinfo
  {pages} {3188} (\bibinfo {year} {1984})}\BibitemShut {NoStop}%
\bibitem [{\citenamefont {Frenkel}\ and\ \citenamefont
  {Smit}(2002)}]{frenkel_understanding_2002}%
  \BibitemOpen
  \bibfield  {author} {\bibinfo {author} {\bibfnamefont {D.}~\bibnamefont
  {Frenkel}}\ and\ \bibinfo {author} {\bibfnamefont {B.}~\bibnamefont {Smit}},\
  }\href@noop {} {\emph {\bibinfo {title} {Understanding {Molecular}
  {Simulation}: {From} {Algorithms} to {Applications}}}},\ \bibinfo {edition}
  {2nd}\ ed.\ (\bibinfo  {publisher} {Academic Press},\ \bibinfo {address} {San
  Diego},\ \bibinfo {year} {2002})\BibitemShut {NoStop}%
\bibitem [{\citenamefont {Chaudhuri}, \citenamefont {Berthier},\ and\
  \citenamefont {Sastry}(2010)}]{chaudhuri_jamming_2010}%
  \BibitemOpen
  \bibfield  {author} {\bibinfo {author} {\bibfnamefont {P.}~\bibnamefont
  {Chaudhuri}}, \bibinfo {author} {\bibfnamefont {L.}~\bibnamefont {Berthier}},
  \ and\ \bibinfo {author} {\bibfnamefont {S.}~\bibnamefont {Sastry}},\ }\href
  {\doibase 10.1103/PhysRevLett.104.165701} {\bibfield  {journal} {\bibinfo
  {journal} {Phys. Rev. Lett.}\ }\textbf {\bibinfo {volume} {104}},\ \bibinfo
  {pages} {165701} (\bibinfo {year} {2010})}\BibitemShut {NoStop}%
\bibitem [{\citenamefont {Parisi}\ and\ \citenamefont
  {Zamponi}(2010)}]{parisi_mean_field_2010}%
  \BibitemOpen
  \bibfield  {author} {\bibinfo {author} {\bibfnamefont {G.}~\bibnamefont
  {Parisi}}\ and\ \bibinfo {author} {\bibfnamefont {F.}~\bibnamefont
  {Zamponi}},\ }\href {\doibase 10.1103/RevModPhys.82.789} {\bibfield
  {journal} {\bibinfo  {journal} {Rev. Mod. Phys.}\ }\textbf {\bibinfo {volume}
  {82}},\ \bibinfo {pages} {789} (\bibinfo {year} {2010})}\BibitemShut
  {NoStop}%
\bibitem [{\citenamefont {Baranau}\ and\ \citenamefont
  {Tallarek}(2014{\natexlab{b}})}]{baranau_jamming_2014}%
  \BibitemOpen
  \bibfield  {author} {\bibinfo {author} {\bibfnamefont {V.}~\bibnamefont
  {Baranau}}\ and\ \bibinfo {author} {\bibfnamefont {U.}~\bibnamefont
  {Tallarek}},\ }\href {\doibase 10.1039/C4SM01439A} {\bibfield  {journal}
  {\bibinfo  {journal} {Soft Matter}\ }\textbf {\bibinfo {volume} {10}},\
  \bibinfo {pages} {7838} (\bibinfo {year} {2014}{\natexlab{b}})}\BibitemShut
  {NoStop}%
\bibitem [{\citenamefont {Baranau}\ and\ \citenamefont
  {Tallarek}(2015)}]{baranau_how_2015}%
  \BibitemOpen
  \bibfield  {author} {\bibinfo {author} {\bibfnamefont {V.}~\bibnamefont
  {Baranau}}\ and\ \bibinfo {author} {\bibfnamefont {U.}~\bibnamefont
  {Tallarek}},\ }\href {\doibase 10.1063/1.4927077} {\bibfield  {journal}
  {\bibinfo  {journal} {J. Chem. Phys.}\ }\textbf {\bibinfo {volume} {143}},\
  \bibinfo {pages} {044501} (\bibinfo {year} {2015})}\BibitemShut {NoStop}%
\bibitem [{\citenamefont {Coslovich}, \citenamefont {Berthier},\ and\
  \citenamefont {Ozawa}(2017)}]{coslovich_exploring_2017}%
  \BibitemOpen
  \bibfield  {author} {\bibinfo {author} {\bibfnamefont {D.}~\bibnamefont
  {Coslovich}}, \bibinfo {author} {\bibfnamefont {L.}~\bibnamefont {Berthier}},
  \ and\ \bibinfo {author} {\bibfnamefont {M.}~\bibnamefont {Ozawa}},\ }\href
  {\doibase 10.21468/SciPostPhys.3.4.027} {\bibfield  {journal} {\bibinfo
  {journal} {SciPost Phys.}\ }\textbf {\bibinfo {volume} {3}},\ \bibinfo
  {pages} {027} (\bibinfo {year} {2017})}\BibitemShut {NoStop}%
\bibitem [{\citenamefont {Aste}\ and\ \citenamefont
  {Coniglio}(2004)}]{aste_cell_2004}%
  \BibitemOpen
  \bibfield  {author} {\bibinfo {author} {\bibfnamefont {T.}~\bibnamefont
  {Aste}}\ and\ \bibinfo {author} {\bibfnamefont {A.}~\bibnamefont
  {Coniglio}},\ }\href {\doibase 10.1209/epl/i2003-10284-x} {\bibfield
  {journal} {\bibinfo  {journal} {Europhys. Lett.}\ }\textbf {\bibinfo {volume}
  {67}},\ \bibinfo {pages} {165} (\bibinfo {year} {2004})}\BibitemShut
  {NoStop}%
\bibitem [{\citenamefont {Parisi}\ and\ \citenamefont
  {Zamponi}(2005)}]{parisi_ideal_2005}%
  \BibitemOpen
  \bibfield  {author} {\bibinfo {author} {\bibfnamefont {G.}~\bibnamefont
  {Parisi}}\ and\ \bibinfo {author} {\bibfnamefont {F.}~\bibnamefont
  {Zamponi}},\ }\href {\doibase 10.1063/1.2041507} {\bibfield  {journal}
  {\bibinfo  {journal} {J. Chem. Phys.}\ }\textbf {\bibinfo {volume} {123}},\
  \bibinfo {pages} {144501} (\bibinfo {year} {2005})}\BibitemShut {NoStop}%
\bibitem [{\citenamefont {Jadrich}\ and\ \citenamefont
  {Schweizer}(2013)}]{jadrich_equilibrium_2013}%
  \BibitemOpen
  \bibfield  {author} {\bibinfo {author} {\bibfnamefont {R.}~\bibnamefont
  {Jadrich}}\ and\ \bibinfo {author} {\bibfnamefont {K.~S.}\ \bibnamefont
  {Schweizer}},\ }\href {\doibase doi:10.1063/1.4816275} {\bibfield  {journal}
  {\bibinfo  {journal} {J. Chem. Phys.}\ }\textbf {\bibinfo {volume} {139}},\
  \bibinfo {pages} {054501} (\bibinfo {year} {2013})}\BibitemShut {NoStop}%
\bibitem [{\citenamefont {Berthier}, \citenamefont {Jacquin},\ and\
  \citenamefont {Zamponi}(2011)}]{berthier_microscopic_2011}%
  \BibitemOpen
  \bibfield  {author} {\bibinfo {author} {\bibfnamefont {L.}~\bibnamefont
  {Berthier}}, \bibinfo {author} {\bibfnamefont {H.}~\bibnamefont {Jacquin}}, \
  and\ \bibinfo {author} {\bibfnamefont {F.}~\bibnamefont {Zamponi}},\ }\href
  {\doibase 10.1103/PhysRevE.84.051103} {\bibfield  {journal} {\bibinfo
  {journal} {Phys. Rev. E}\ }\textbf {\bibinfo {volume} {84}},\ \bibinfo
  {pages} {051103} (\bibinfo {year} {2011})}\BibitemShut {NoStop}%
\bibitem [{\citenamefont {Kirkwood}(1950)}]{kirkwood_critique_1950}%
  \BibitemOpen
  \bibfield  {author} {\bibinfo {author} {\bibfnamefont {J.~G.}\ \bibnamefont
  {Kirkwood}},\ }\href {\doibase doi:10.1063/1.1747635} {\bibfield  {journal}
  {\bibinfo  {journal} {J. Chem. Phys.}\ }\textbf {\bibinfo {volume} {18}},\
  \bibinfo {pages} {380} (\bibinfo {year} {1950})}\BibitemShut {NoStop}%
\bibitem [{\citenamefont {Buehler}\ \emph {et~al.}(1951)\citenamefont
  {Buehler}, \citenamefont {Wentorf~Jr.}, \citenamefont {Hirschfelder},\ and\
  \citenamefont {Curtiss}}]{buehler_free_1951}%
  \BibitemOpen
  \bibfield  {author} {\bibinfo {author} {\bibfnamefont {R.~J.}\ \bibnamefont
  {Buehler}}, \bibinfo {author} {\bibfnamefont {R.~H.}\ \bibnamefont
  {Wentorf~Jr.}}, \bibinfo {author} {\bibfnamefont {J.~O.}\ \bibnamefont
  {Hirschfelder}}, \ and\ \bibinfo {author} {\bibfnamefont {C.~F.}\
  \bibnamefont {Curtiss}},\ }\href {\doibase 10.1063/1.1747991} {\bibfield
  {journal} {\bibinfo  {journal} {J. Chem. Phys.}\ }\textbf {\bibinfo {volume}
  {19}},\ \bibinfo {pages} {61} (\bibinfo {year} {1951})}\BibitemShut {NoStop}%
\bibitem [{\citenamefont {Wood}(1952)}]{wood_note_1952}%
  \BibitemOpen
  \bibfield  {author} {\bibinfo {author} {\bibfnamefont {W.~W.}\ \bibnamefont
  {Wood}},\ }\href {\doibase doi:10.1063/1.1700747} {\bibfield  {journal}
  {\bibinfo  {journal} {J. Chem. Phys.}\ }\textbf {\bibinfo {volume} {20}},\
  \bibinfo {pages} {1334} (\bibinfo {year} {1952})}\BibitemShut {NoStop}%
\bibitem [{\citenamefont {Salsburg}\ and\ \citenamefont
  {Wood}(1962)}]{salsburg_equation_1962}%
  \BibitemOpen
  \bibfield  {author} {\bibinfo {author} {\bibfnamefont {Z.~W.}\ \bibnamefont
  {Salsburg}}\ and\ \bibinfo {author} {\bibfnamefont {W.~W.}\ \bibnamefont
  {Wood}},\ }\href {\doibase doi:10.1063/1.1733163} {\bibfield  {journal}
  {\bibinfo  {journal} {J. Chem. Phys.}\ }\textbf {\bibinfo {volume} {37}},\
  \bibinfo {pages} {798} (\bibinfo {year} {1962})}\BibitemShut {NoStop}%
\bibitem [{\citenamefont {Masri}\ \emph {et~al.}(2009)\citenamefont {Masri},
  \citenamefont {Brambilla}, \citenamefont {Pierno}, \citenamefont {Petekidis},
  \citenamefont {Schofield}, \citenamefont {Berthier},\ and\ \citenamefont
  {Cipelletti}}]{masri_dynamic_2009}%
  \BibitemOpen
  \bibfield  {author} {\bibinfo {author} {\bibfnamefont {D.~E.}\ \bibnamefont
  {Masri}}, \bibinfo {author} {\bibfnamefont {G.}~\bibnamefont {Brambilla}},
  \bibinfo {author} {\bibfnamefont {M.}~\bibnamefont {Pierno}}, \bibinfo
  {author} {\bibfnamefont {G.}~\bibnamefont {Petekidis}}, \bibinfo {author}
  {\bibfnamefont {A.~B.}\ \bibnamefont {Schofield}}, \bibinfo {author}
  {\bibfnamefont {L.}~\bibnamefont {Berthier}}, \ and\ \bibinfo {author}
  {\bibfnamefont {L.}~\bibnamefont {Cipelletti}},\ }\href {\doibase
  10.1088/1742-5468/2009/07/P07015} {\bibfield  {journal} {\bibinfo  {journal}
  {J. Stat. Mech.}\ }\textbf {\bibinfo {volume} {2009}},\ \bibinfo {pages}
  {P07015} (\bibinfo {year} {2009})}\BibitemShut {NoStop}%
\bibitem [{\citenamefont {Brambilla}\ \emph {et~al.}(2009)\citenamefont
  {Brambilla}, \citenamefont {El~Masri}, \citenamefont {Pierno}, \citenamefont
  {Berthier}, \citenamefont {Cipelletti}, \citenamefont {Petekidis},\ and\
  \citenamefont {Schofield}}]{brambilla_probing_2009}%
  \BibitemOpen
  \bibfield  {author} {\bibinfo {author} {\bibfnamefont {G.}~\bibnamefont
  {Brambilla}}, \bibinfo {author} {\bibfnamefont {D.}~\bibnamefont {El~Masri}},
  \bibinfo {author} {\bibfnamefont {M.}~\bibnamefont {Pierno}}, \bibinfo
  {author} {\bibfnamefont {L.}~\bibnamefont {Berthier}}, \bibinfo {author}
  {\bibfnamefont {L.}~\bibnamefont {Cipelletti}}, \bibinfo {author}
  {\bibfnamefont {G.}~\bibnamefont {Petekidis}}, \ and\ \bibinfo {author}
  {\bibfnamefont {A.~B.}\ \bibnamefont {Schofield}},\ }\href {\doibase
  10.1103/PhysRevLett.102.085703} {\bibfield  {journal} {\bibinfo  {journal}
  {Phys. Rev. Lett.}\ }\textbf {\bibinfo {volume} {102}},\ \bibinfo {pages}
  {085703} (\bibinfo {year} {2009})}\BibitemShut {NoStop}%
\bibitem [{\citenamefont {P\'erez-\'Angel}\ \emph {et~al.}(2011)\citenamefont
  {P\'erez-\'Angel}, \citenamefont {S\'anchez-D\'iaz}, \citenamefont
  {Ram\'irez-Gonz\'alez}, \citenamefont {Ju\'arez-Maldonado}, \citenamefont
  {Vizcarra-Rend\'on},\ and\ \citenamefont
  {Medina-Noyola}}]{perez_angel_equilibration_2011}%
  \BibitemOpen
  \bibfield  {author} {\bibinfo {author} {\bibfnamefont {G.}~\bibnamefont
  {P\'erez-\'Angel}}, \bibinfo {author} {\bibfnamefont {L.~E.}\ \bibnamefont
  {S\'anchez-D\'iaz}}, \bibinfo {author} {\bibfnamefont {P.~E.}\ \bibnamefont
  {Ram\'irez-Gonz\'alez}}, \bibinfo {author} {\bibfnamefont {R.}~\bibnamefont
  {Ju\'arez-Maldonado}}, \bibinfo {author} {\bibfnamefont {A.}~\bibnamefont
  {Vizcarra-Rend\'on}}, \ and\ \bibinfo {author} {\bibfnamefont
  {M.}~\bibnamefont {Medina-Noyola}},\ }\href {\doibase
  10.1103/PhysRevE.83.060501} {\bibfield  {journal} {\bibinfo  {journal} {Phys.
  Rev. E}\ }\textbf {\bibinfo {volume} {83}},\ \bibinfo {pages} {060501}
  (\bibinfo {year} {2011})}\BibitemShut {NoStop}%
\bibitem [{\citenamefont {Zaccarelli}, \citenamefont {Liddle},\ and\
  \citenamefont {Poon}(2014)}]{zaccarelli_polydispersity_2014}%
  \BibitemOpen
  \bibfield  {author} {\bibinfo {author} {\bibfnamefont {E.}~\bibnamefont
  {Zaccarelli}}, \bibinfo {author} {\bibfnamefont {S.~M.}\ \bibnamefont
  {Liddle}}, \ and\ \bibinfo {author} {\bibfnamefont {W.~C.~K.}\ \bibnamefont
  {Poon}},\ }\href {\doibase 10.1039/C4SM02321H} {\bibfield  {journal}
  {\bibinfo  {journal} {Soft Matter}\ }\textbf {\bibinfo {volume} {11}},\
  \bibinfo {pages} {324} (\bibinfo {year} {2014})}\BibitemShut {NoStop}%
\bibitem [{\citenamefont {Edwards}\ and\ \citenamefont
  {Oakeshott}(1989)}]{edwards_theory_1989}%
  \BibitemOpen
  \bibfield  {author} {\bibinfo {author} {\bibfnamefont {S.~F.}\ \bibnamefont
  {Edwards}}\ and\ \bibinfo {author} {\bibfnamefont {R.~B.~S.}\ \bibnamefont
  {Oakeshott}},\ }\href@noop {} {\bibfield  {journal} {\bibinfo  {journal}
  {Physica A}\ }\textbf {\bibinfo {volume} {157}},\ \bibinfo {pages} {1080}
  (\bibinfo {year} {1989})}\BibitemShut {NoStop}%
\bibitem [{\citenamefont {Bowles}\ and\ \citenamefont
  {Ashwin}(2011)}]{bowles_edwards_2011}%
  \BibitemOpen
  \bibfield  {author} {\bibinfo {author} {\bibfnamefont {R.~K.}\ \bibnamefont
  {Bowles}}\ and\ \bibinfo {author} {\bibfnamefont {S.~S.}\ \bibnamefont
  {Ashwin}},\ }\href {\doibase 10.1103/PhysRevE.83.031302} {\bibfield
  {journal} {\bibinfo  {journal} {Phys. Rev. E}\ }\textbf {\bibinfo {volume}
  {83}},\ \bibinfo {pages} {031302} (\bibinfo {year} {2011})}\BibitemShut
  {NoStop}%
\bibitem [{\citenamefont {Bi}\ \emph {et~al.}(2015)\citenamefont {Bi},
  \citenamefont {Henkes}, \citenamefont {Daniels},\ and\ \citenamefont
  {Chakraborty}}]{bi_statistical_2015}%
  \BibitemOpen
  \bibfield  {author} {\bibinfo {author} {\bibfnamefont {D.}~\bibnamefont
  {Bi}}, \bibinfo {author} {\bibfnamefont {S.}~\bibnamefont {Henkes}}, \bibinfo
  {author} {\bibfnamefont {K.~E.}\ \bibnamefont {Daniels}}, \ and\ \bibinfo
  {author} {\bibfnamefont {B.}~\bibnamefont {Chakraborty}},\ }\href {\doibase
  10.1146/annurev-conmatphys-031214-014336} {\bibfield  {journal} {\bibinfo
  {journal} {Annu. Rev. Condens. Matter Phys.}\ }\textbf {\bibinfo {volume}
  {6}},\ \bibinfo {pages} {63} (\bibinfo {year} {2015})}\BibitemShut {NoStop}%
\bibitem [{\citenamefont {Torquato}\ and\ \citenamefont
  {Stillinger}(2010)}]{torquato_jammed_2010}%
  \BibitemOpen
  \bibfield  {author} {\bibinfo {author} {\bibfnamefont {S.}~\bibnamefont
  {Torquato}}\ and\ \bibinfo {author} {\bibfnamefont {F.~H.}\ \bibnamefont
  {Stillinger}},\ }\href {\doibase 10.1103/RevModPhys.82.2633} {\bibfield
  {journal} {\bibinfo  {journal} {Rev. Mod. Phys.}\ }\textbf {\bibinfo {volume}
  {82}},\ \bibinfo {pages} {2633} (\bibinfo {year} {2010})}\BibitemShut
  {NoStop}%
\end{thebibliography}%

\end{document}